\documentclass[twocolumn,prl,superscriptaddress]{revtex4-1}

\usepackage{amsmath,graphicx,xcolor,hyperref}
\usepackage[normalem]{ulem}

\providecommand{\vect}[1]{\bf{#1}}
\providecommand{\pedex}[1]{\ensuremath{_\text{#1}}}
\providecommand{\abs}[1]{\ensuremath{\bigl \lvert #1 \bigr \rvert}}

\begin{document}

%\title{Ultrafast carrier relaxation and polaronic mass enhancement \\
%in the hybrid halide perovskite CH$_3$NH$_3$PbI$_3$ \\ from multi-phonon Fr\"ohlich coupling}

\title{Carrier {Lifetimes} and {Polaronic Mass Enhancement }
in the {Hybrid}\\[3pt] {Halide Perovskite} CH$_3$NH$_3$PbI$_3$ from {Multiphonon} Fr\"ohlich {Coupling}}

\author{Martin Schlipf}
\affiliation{Department of Materials, University of Oxford, Parks Road, Oxford OX1 3PH, United Kingdom}
\author{Samuel Ponc\'e}
\affiliation{Department of Materials, University of Oxford, Parks Road, Oxford OX1 3PH, United Kingdom}
\author{Feliciano Giustino}
\email{feliciano.giustino@materials.ox.ac.uk}
\affiliation{Department of Materials, University of Oxford, Parks Road, Oxford OX1 3PH, United Kingdom}
\affiliation{Department of Materials Science and Engineering, Cornell University, Ithaca, New York 14853, USA}

\begin{abstract}
We elucidate the nature of the electron-phonon interaction {in the archetypal hybrid perovskite CH$_3$NH$_3$PbI$_3$}
using \emph{ab initio} many-body calculations and an {exactly solvable} model.
We demonstrate that electrons 
and holes near the band edges primarily interact with three distinct groups of longitudinal-optical 
vibrations, in order of importance: 
the stretching of the {Pb--I} bond, the {bending} of the {Pb--I--Pb} bonds, and the libration of 
the organic cations. These polar phonons induce ultrafast intraband carrier 
relaxation over {timescales} of {$6-30$}~fs and yield polaron effective masses 
28\% heavier than the bare band masses. These findings allow us to rationalize previous 
experimental observations and provide a key to understanding carrier dynamics in halide perovskites.

\end{abstract}

\maketitle

Recently, hybrid organic-inorganic lead-halide perovskites {like} CH$_3$NH$_3$PbI$_3$
emerged as promising materials for high-performance solar cells~\cite{lee12,kim12}.
These systems are unique semiconductors, insofar {as} solution-processed 
thin films exhibit
optoelectronic properties on par with monocrystalline inorganic semiconductors~\cite{Yablonovitch2012, Green2018}. 
This exceptional performance originates from the direct gap 
near the Schockley-Queisser limit, the low and balanced carrier effective masses, the long recombination 
lifetimes, the tolerance to defects, and the comparatively high carrier mobilities for solution-processed 
semiconductors~\cite{ghs14, Johnston2016, Herz2017}. Several of these 
attributes are connected with the electron-phonon interaction (EPI)~\cite{wehr14,Wehrenfennig2014,kara15,milot15,
herz16,Filippetti2016}: the blueshift of the optical absorption onset and the photoluminescence peak with 
temperature~\cite{Wu2014, milot15, dar16}; presumably the long recombination lifetimes~\cite{ztlr15, kepe15,emd16} 
(combined with the Rashba-Dresselhaus effect, EPIs may induce spin-forbidden transitions near the band extrema);
and a hot-phonon bottleneck possibly originating from an imbalance between electron 
thermalization via optical phonon emission and heat transport~\cite{Price2015,Yang2016, Yang2017b}. 

The nature of EPIs in CH$_3$NH$_3$PbI$_3$ is being intensely debated. Transient photoluminescence 
studies indicate that the EPI in this compound is dominated by the Fr\"ohlich coupling of the carriers 
with {one} {longitudinal-optical} (LO) phonon \cite{wrig16} identified by infrared spectroscopy 
\cite{perez15}. Yet some studies suggest acoustic phonons playing an 
important role~\cite{kara15,Even2016} and limiting charge transport
via scattering~\cite{Zhu2015, Chen2016}.
{Recent theoretical modeling debated whether LO~\cite{Frost2017a} or acoustic~\cite{Zhang2017} 
phonon scattering limits the carrier mobility.}
By offering an atomic-scale 
perspective on the role of each phonon, {\it ab initio} calculations can contribute to elucidating 
the fundamental mechanisms underpinning EPIs in CH$_3$NH$_3$PbI$_3$.

In this work, we study the EPIs from first principles by employing state-of-the-art 
many-body calculations. We demonstrate that {CH$_3$NH$_3$PbI$_3$ is unique among polar
semiconductors in that not a single but} three distinct groups of LO phonons dominate the EPI:
two associated with the PbI$_6$ octahedra and one with the librations 
of the organic cations, while acoustic phonons {hardly} contribute. 
This {unusual} {multiphonon}
Fr\"ohlich coupling is {a direct cause of the structural complexity of hybrid perovskites}
and {is} responsible for the ultrafast relaxation 
of photoexcited carriers and for a moderate polaronic enhancement of {their} masses.

We perform {\emph{ab initio}} calculations within the low-temperature orthorhombic $Pnma$ phase~\cite{baik13} of CH$_3$NH$_3$PbI$_3$
shown in Supplemental {Material} Fig.~S1~\cite{suppl}.
\nocite{perez15,QE2017,wannier90,epw,ca80,pz81,hama13,sg15,Lejaeghere2016,Setten2018,gcl07,
rycr09,fvg15,vg15,Sjakste2015,Marronnier2017,kara15,wrig16,Perez2017,Seeger2004,
Grimvall1981,Galkowski2016,froe54}
{These calculations} are unusually challenging {due to} the necessity of including spin-orbit coupling~\cite{epjk13} {and} quasiparticle $GW$ corrections~\cite{umd14,fg14}{,}
{the} large unit cell consisting of 48 atoms{,} and the sensitivity 
of EPI calculations to the Brillouin-zone integration. Various approximations have been employed in order 
to circumvent some of these obstacles: {\citet{Kawai2015}} calculated EPIs in the 
simpler model system CsPbI$_3$ by neglecting spin-orbit effects. In Refs.~\onlinecite{spm16,Bokdam2016},
the authors investigated the cubic, 12-atom unit cell, aligning the organic cations in a ferroelectric 
configuration. In Ref.~\onlinecite{wrig16}, the authors studied the orthorhombic phase considering only the {long-range} coupling to {polar phonons}.

Here we go beyond these previous studies {by calculating~\cite{suppl}
and analyzing the complete \emph{ab initio} self-energy arising from EPIs in
many-body perturbation theory. Since CH$_3$NH$_3$PbI$_3$~is in the {weak-coupling}
regime~\cite{weak-regime},\nocite{Mahan2000, Devreese2009} we use perturbation theory within
the retarded Fan-Migdal electron self-energy $\Sigma_{n{\bf k}}(\varepsilon)$~%
\cite{fan51,Migdal1958,gius17}}:
  \begin{multline} \label{eq:self-energy}
  \Sigma_{n{\bf k}}(\varepsilon) = \sum_{m\nu} \int \!\frac{d{\bf q}}{\Omega\pedex{BZ}}
  \abs{g_{mn\nu}({\bf k}, {\bf q})}^2  
  \\ \times
  \sum_{\pm} \frac{n_{\vect q\nu} + (1 \pm [2 f_{m\vect k + \vect q} - 1]) / 2}
  {\varepsilon - \varepsilon_{m\vect k + \vect q} \pm \hbar \omega_{\vect q\nu} + i \eta}.
  \end{multline}
Here $\varepsilon_{n\vect k}$ denotes the Kohn-Sham eigenvalue for the band $n$ and {wave vector} ${\bf k}$,
$\omega_{\vect q\nu}$ is the vibrational frequency for the {phonon branch} $\nu$ and {wave vector} ${\bf q}$. 
$g_{mn\nu}({\bf k}, {\bf q})$ are the electron-phonon matrix elements, and the integral is performed over the Brillouin 
zone of volume $\Omega\pedex{BZ}$. The temperature is included via the Fermi-Dirac 
$f_{m\vect k + \vect q}$ and Bose-Einstein $n_{\vect q\nu}$
occupations. Using the self-energy in 
Eq.~(\ref{eq:self-energy}), we determine the quasiparticle lifetimes and effective mass 
renormalization within Brillouin-Wigner perturbation theory
as $E_{n{\bf k}}+i\Gamma_{n{\bf k}} 
= \varepsilon_{n{\bf k}} + Z_{n\vect k} \Sigma_{n{\bf k}}(\varepsilon_{n \vect k})$
with $Z_{n\vect k} = [1 - \left.\text{Re}(\partial \Sigma_{{n\vect k}} / \partial {\varepsilon}
)\right|_{\varepsilon_{n\vect k}}]^{-1}$, where
$E_{n{\bf k}}$ is the quasiparticle and $\Gamma_{n{\bf k}}$ its associated broadening~\cite{Hedin1969,gius17}.
The conventional Rayleigh-Schr\"odinger approach is obtained by setting $Z_{n\vect k} = 1$.
The quasiparticle 
lifetime is obtained via $\tau_{n{\bf k}} = \hbar/2\Gamma_{n{\bf k}}$~\cite{Hedin1969,gius17}.
The renormalization of the effective masses is calculated from the {${\vect k}$ derivatives}
of $E_{n{\bf k}}$~\cite{suppl}.
The quasiparticle mass can be expressed in terms of the bare band mass as 
$m^{{\rm QP},*}_{n{\bf k}} = (1+\lambda_{n{\bf k}})m^*_{n{\bf k}}$, where $\lambda_{n{\bf k}}$ is 
analogous to the mass-enhancement parameter in metals~\cite{Grimvall1981}.
%
% we are also neglecting the variation of Z_k with k
%
% Note that the convergengce of the real part of Sigma wrt empty states and the Debye-Waller effect
% can be offset by resetting the zero of the energy axis at the band extrema. However this change
% only applies to the band bottom or top, not the other states (there must be only one shift for
% all wavevectors). 
%

\begin{figure}
  \includegraphics{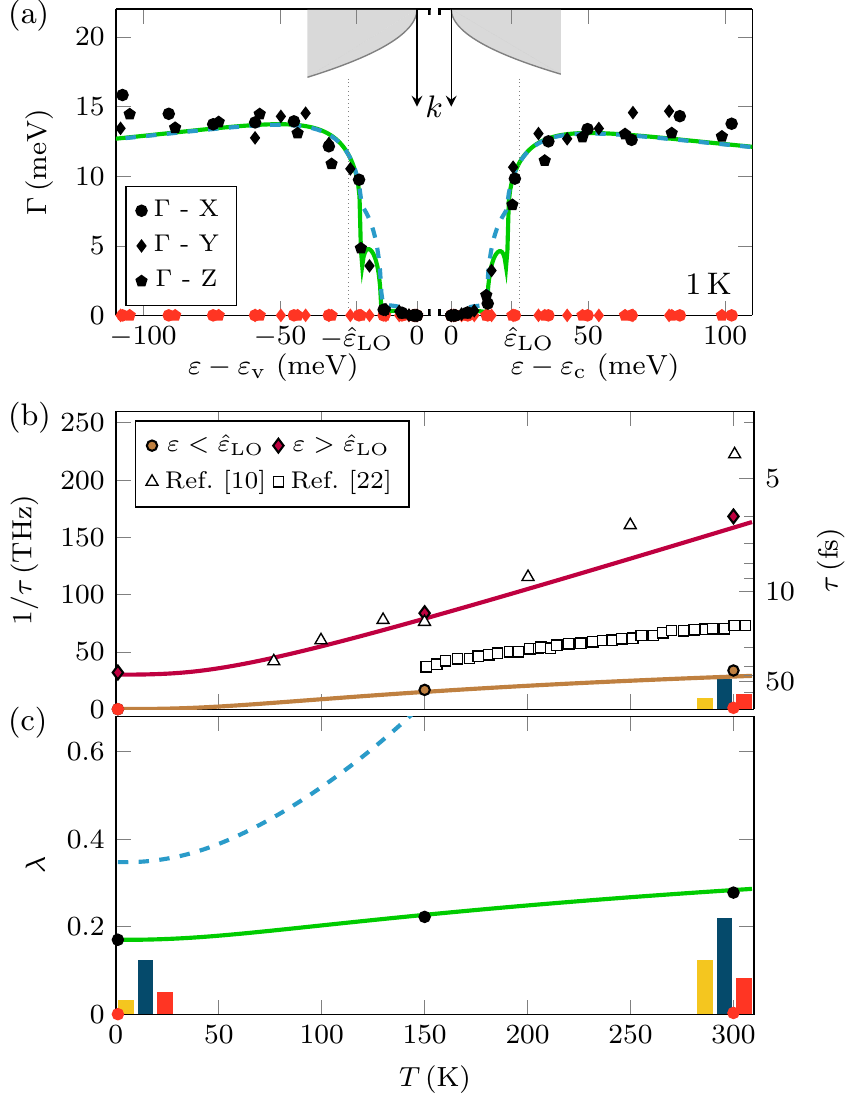}
  \caption{
  (a) Quasiparticle broadening in CH$_3$NH$_3$PbI$_3$ from EPIs near the band edges at 1~K.
  The carrier energy for {wave vectors} along various high-symmetry directions
  is {referred} to the band edge, and the bands are indicated schematically.
  {{\scshape epw} calculations are shown as filled black symbols; red symbols indicate the contribution of all
  phonons with energy $\lesssim 2$~meV; the lines are calculated with Eq.~\eqref{eq:model}
   using {Rayleigh-Schr\"odinger} ({teal,} dashed {line}) or Brillouin-Wigner ({green}, solid) perturbation theory.}
  (b) Smallest relaxation rates ({left-hand} axis) and largest lifetime ({right-hand} axis)
  for holes with energy above ({purple} diamond and line) or below
  ({brown} circle and line) the maximum energy of polar
  phonons $\hat\varepsilon\pedex{LO} \sim 22~\text{meV}$.
  {Filled symbols are {\scshape epw} calculations, lines are from the {multiphonon} model.}
  The data for electrons are practically indistinguishable from the holes and
  are not shown.
  The open symbols are {the relaxation rates measured by}
  Ref.~\citenum{kara15} (triangle) and Ref.~\citenum{wrig16} (square). (c)
  Mass enhancement parameter for holes {(data for electrons are practically indistinguishable)}.
  {\scshape epw} calculations are shown
  as filled symbols. The lines are from {Rayleigh-Schr\"odinger} ({teal,} dashed {line}) 
  {or Brillouin-Wigner} ({green}, solid {line}) perturbation theory.
  The colored bars indicate the contributions of
  the three groups of LO modes shown in Fig.~\ref{fig2}, using the same color code.
  }
  \label{fig1}
\end{figure}
  
%\section{Results}
%
%\subsection{\emph{Ab initio} carrier lifetimes and mass enhancement}
Figure~\ref{fig1}(a) shows the calculated quasiparticle broadening near the valence and conduction 
band edges at 1~K. For completeness, the band structure and phonon dispersion are shown in
Supplemental {Material} Fig.~S2~\cite{suppl}. A similar trend is found for a range of 
temperatures.  The broadening is essentially the same along different high-symmetry 
lines, indicating that the EPI in CH$_3$NH$_3$PbI$_3$ is isotropic. At the band edges,
{we have sharp quasiparticle bands with vanishing broadening}. 
Here the absorption and emission of phonons are 
forbidden, because there are no thermally excited phonons to be absorbed and 
phonon emission would send the carriers inside the band gap. Away from the band 
edges, the density of final electron or hole states available for scattering increases, and we see a 
finite broadening. The {red} symbols in Fig.~\ref{fig1}(a) show the contribution
arising from acoustic phonons and from low-energy transverse-optical (TO) phonons 
($\hbar\omega \lesssim 2$~meV).
{Their vanishingly small contributions indicate} that the EPI is
dominated by polar modes (see Supplemental {Material} Fig.~S3 for the same plot at 
300~K~\cite{suppl}). The {steplike} feature seen in 
Fig.~\ref{fig1}(a) arises primarily from LO phonons: the steep increase is due to the 
activation of multiple electron-LO phonon scattering channels at increasingly higher energy,
while %the 
%plateau relates to the fact that in the case of Fr\"ohlich interactions
the quasiparticle broadening is determined by the band 
velocity rather than the density of states as in metals~\cite{vg15},
{leading to the plateau at higher energies}. 
The characteristic energy 
scale of this step is set by the most energetic LO phonon around 22~meV, 
as we discuss below. 
The quasiparticle broadening at the plateau corresponds to an intrinsic, carrier lifetime 
for electrons and holes of approximately 30~fs at 1~K and 6~fs at 300~K.

Figure~\ref{fig1}(b) shows that the lifetimes decrease rapidly with increasing temperature.
{Our calculations explain the differences between the experimental photoconductivity (PC) data
of Ref.~\onlinecite{kara15} and the photoluminescence (PL) linewidth of Ref.~\onlinecite{wrig16}.
In the PL experiment, carriers recombine close to the band edges and experience
a smaller EPI than carriers above the most energetic LO phonon at 22~meV
probed by the PC experiment.
In fact, the temperature dependence of the PC scattering rate 
follows a $T^{-3/2}$ law; precisely what we observe if we
calculate the average scattering time~(see {Supplemental Material} Fig.~S3~\cite{suppl}).
These observations highlight
the importance of understanding the energy distribution of the
charge carriers probed by different experiments, which report scattering times between {4 and 15}~fs
at 300~K.
{The mismatch between measured and calculated phonon frequency~\cite{Perez2017} may 
lead to up to 25\% smaller scattering times~(see {Supplemental Material} Figure~S4~\cite{suppl}).}
The {femtosecond timescale} of the EPI is} too small for optical phonons
to contribute to the hot-phonon bottleneck, which takes place on {picosecond} timescales. Therefore our
calculations support
the recent proposal of Ref.~\onlinecite{Yang2017b} that the bottleneck must relate to the 
up-conversion of acoustic phonons.

Figure~\ref{fig1}(c) shows the calculated electron-phonon renormalization of the
electron and hole masses of CH$_3$NH$_3$PbI$_3$.
At 300~K, the EPI enhances both electron and hole masses by 28\%.
Importantly the effective mass is mostly independent of temperature, increasing by only 9\% 
from 1 to 300~K.
This result is {counterintuitive}, as electron-phonon effects are usually expected to 
become more pronounced as more phonons become available to dress the carriers. What happens here is that 
the increase of the quasiparticle broadening with temperature offsets the increase in the
real part of the {self-energy}. This {nontrivial} effect mirrors an analogous mechanism
discussed for simple metals in Ref.~\citenum{Grimvall1968}.
Our calculations are in agreement with {magnetotransport} measurements indicating a reduced 
exciton effective mass independent of temperature for CH$_3$NH$_3$PbI$_3$~\cite{Miyata2015}, and increasing 
by {$5\%-13\%$} from 2 to 300~K in related halide perovskites~\cite{Galkowski2016}. Our calculations are 
consistent with the fact that accurate calculations of the band effective masses using the $GW$
method agree with experiments within {$10\%-20\%$}~\cite{umd14,fvg15}.
The results based on the less accurate 
Rayleigh-Schr\"odinger perturbation theory (dashed lines) significantly 
overestimate the mass renormalization and do not agree with experimental evidence. {Therefore}, 
when discussing EPI in halide perovskites, it is important to use the more accurate Brillouin-Wigner  method.

\begin{figure}[b]
  \includegraphics{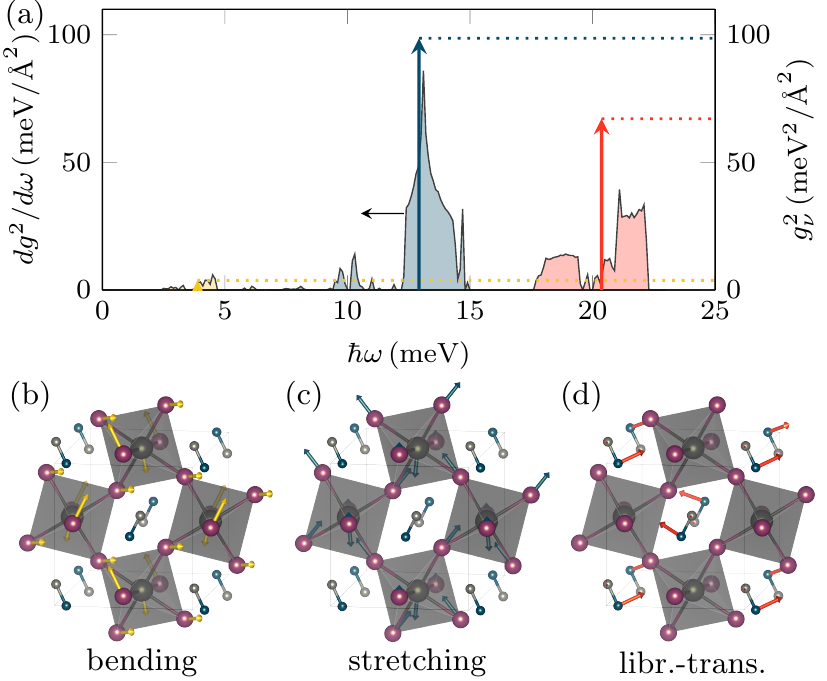}
  \caption{
  (a) Density of electron-phonon coupling strength associated with polar phonons in CH$_3$NH$_3$PbI$_3$,
  $g^2(\hbar\omega)$, as defined in the Supplemental {Material}~\cite{suppl}. The yellow, blue,
  and red regions correspond to {bending}, stretching, and libration-translation modes, respectively. The vertical arrows
  indicate the energies and couplings of the compact, three-phonon model used in the analysis. (b){--}(d)
  Schematic ball-and-stick representations of the three groups of vibrations appearing in~(a).
  {The Pb and I atoms are at the centers and at the corners of the octahedra, respectively,
  C is in {gray} and N is in blue. The H atoms are not shown for clarity.}
  }
  \label{fig2}
\end{figure}
In order to rationalize the above results we analyze the dominant contributions to the EPI self-energy. 
Since CH$_3$NH$_3$PbI$_3$ has 144~phonon branches, it is useful to introduce a simplified model. To this 
aim, we integrate Eq.~\eqref{eq:self-energy} analytically by considering parabolic bands and neglecting 
acoustic and TO phonons.
We consider therefore {a} {Fr\"ohlich model involving multiple
vibrational modes (\emph{multi-phonon Fr\"ohlich model})} with matrix elements 
$g_{mn\nu}({\bf k}, {\bf q}) = \delta_{mn} \,g_\nu/|{\bf q}|$~\cite{froe54}; after using the
residue theorem to evaluate the integral~\cite{suppl}, Eq.~(\ref{eq:self-energy}) becomes
  \begin{equation} \label{eq:model}
  \Sigma_{n\vect k}(\varepsilon) 
  = -\sum_{\nu,\pm} \frac{n_\nu + (1 \pm \sigma_{n})/2}{\Omega\pedex{BZ}\varepsilon_{n\vect k} / (2\pi^2 g_\nu^2\, k)} \arcsin\!\left[1\!-\!\frac{\varepsilon \pm \hbar \omega_\nu}{\varepsilon_{n\vect k}}\right]^{
\!-\frac{1}{2}}\hspace{-10pt}.
  \end{equation}
where $\sigma_n=\pm 1$ for valence and conduction, respectively.
The EPI strength $g_\nu^2$ appearing in this expression is obtained by averaging the {\it ab initio} 
Fr\"ohlich matrix elements \cite{vg15} over a small sphere around ${\bf q}$~=~0~\cite{suppl}. 
The distribution of coupling strengths is shown in Fig.~\ref{fig2}(a) as a function of phonon energy,
in analogy with the standard Eliashberg function. We see that the Fr\"ohlich interaction is most 
pronounced for three distinct groups of LO phonons. Using the mode analysis of Ref.~\citenum{perez15}, 
we assign these features to the {bending} motion of the {Pb--I--Pb} bonds (yellow), the stretching
motion of the {Pb--I} bonds (blue), and the librations of the organic cations (red).
   Representative 
atomic displacements for these modes are shown in {Figs.~\ref{fig2}(b)--(d)}, respectively. It is useful 
to condense the information presented in Fig.~\ref{fig2}(a) into three {``effective''} phonons carrying the 
same total EPI strength. By averaging these results we obtain~\cite{suppl}
$\hbar\omega_{\rm B} = 3.9~\text{meV}$, $g_{\rm B}^2 = 3.9~\text{meV}^2 \text{\AA}^{-2}$;
$\hbar\omega_{\rm S} = 13.0~\text{meV}$, $g_{\rm S}^2 = 98.9~\text{meV}^2\text{\AA}^{-2}$;
$\hbar\omega_{\rm L} = 20.4~\text{meV}$, $g_{\rm L}^2 = 67.2~\text{meV}^2 \text{\AA}^{-2}$
(the subscripts stand for {bending}, stretching, and libration, respectively). 

The coupling of electrons 
to multiple polar phonons, 
may explain why previous attempts at fitting experimental data using the Fr\"ohlich model yielded LO 
phonon energies ranging from 11.5~\cite{wrig16} to 23.5~meV~\cite{Wu2014}. Furthermore, the multi-LO
coupling offers a possible explanation for the observation of two distinct stages in the relaxation
of hot electrons~\cite{Price2015}.

%{\newcommand{\head}{Mode analysis and spectral function}
%\subsection{\protect\head}
%}
Figure~\ref{fig1}(a) shows a comparison between our model self-energy from Eq.~(\ref{eq:model}) (solid line) 
and the {\emph{ab initio} calculation with {\scshape epw}}. The model is very accurate near the band edges, and starts 
deviating from the {\scshape epw} data for carrier energies {$>60$}~meV from the edges, cf. 
Supplemental {Material} Fig.~S3~\cite{suppl}. This deviation arises from 
the assumption of parabolic bands employed in the model, that progressively breaks down as we move away 
from the band edges. In fact, if we recalculate
the self-energy using the {\emph{first-principles}} density of states, we reproduce the {\it ab initio} data
very accurately, as shown in Supplemental Fig.~{S5}~\cite{suppl}.
Similar comparisons are shown in {Figs.~\ref{fig1}(b) and \ref{fig1}(c)} for the lifetimes and mass 
enhancement. Also in these cases the model captures the essential features of the {\it ab initio} calculations.

The vertical bars in Fig.~\ref{fig1} show a decomposition of
scattering rates and mass enhancement parameter in {the contributions associated with each} of the three polar modes.
At low temperature, the dominant contributions arise from stretching and libration,
while {the bending} modes are only weakly coupled to electrons.
With increasing temperature the {bending} modes become more important due to their
larger Bose-Einstein occupation factor.
{In the orthorhombic structure the CH$_3$NH$_3^+$ cations vibrate 
around their equilibrium sites, but do not spin around the C-N axis as in the tetragonal phase 
(between 160 and 330~K)~\cite{Weller2015,Ren2016,Whitfield2016}.
In the high-temperature cubic phase (above 330~K)~\cite{Weller2015,Ren2016,Whitfield2016}
}the organic cations are fully disordered {and} we expect that 
the librational modes will not contribute to the EPI.
{
To assess the validity of the calculated electron-phonon scattering rates
across a wider temperature range, we simulate orientational disorder of the
organic cations by computing the EPI of CsPbI$_3$ in its cubic phase.
In this case the scattering is reduced by about {$10\%-20\%$} as compared
to the orthorhombic phase of CH$_3$NH$_3$PbI$_3$ due to the absence of librational modes{;
however,}
the energy dependence of the scattering rate is qualitatively similar
to what we obtained for CH$_3$NH$_3$PbI$_3$~(see {Supplemental Material} Fig.~S3~\cite{suppl}).}

%Given the nonlinearity of the BW perturbation theory, switching off
%all but the stretching mode would yield almost the same scattering rate and about 80\% of the mass enhancement
%at room temperature. Analogously, using exclusively the bending mode results in a third of the scattering rate
%and two thirds of the mass enhancement.

Since the model self-energy in Eq.~(\ref{eq:model}) captures the main trends of our complete {\it ab initio} 
calculations, we can use this model to estimate the change in the electron lifetimes and mass enhancement 
resulting from quasiparticle $GW$ corrections to the band structures. \citet{fvg15} showed
that $GW$ calculations increase the electron and hole effective masses of CH$_3$NH$_3$PbI$_3$ 
by {90\% and 80\%}, respectively. At 300~K, using the $GW$ effective masses for
the calculation of the self-energy in Eq.~(\ref{eq:model}) has the effect of
decreasing the relaxation times
at the band edges by about 7\% and of increasing the mass enhancement factor 
by 10\% as compared to the values shown in Fig.~\ref{fig1}. 
This indicates that $GW$ corrections only induce small quantitative changes 
to the picture, but do not change the essence of the present analysis.

\begin{figure}[b]
  \includegraphics{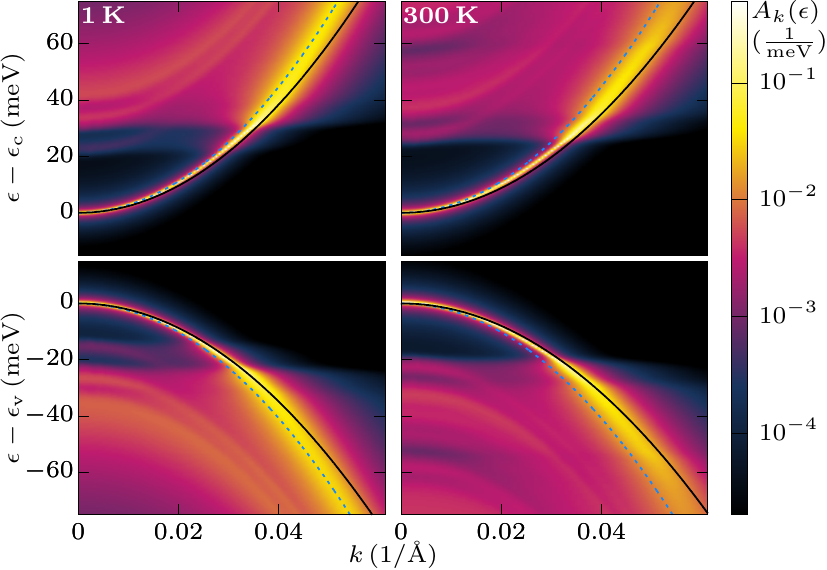}
  \caption{
  Calculated spectral function of CH$_3$NH$_3$PbI$_3$ including EPIs at 1 and 300~K near
  the {edges} of conduction (c) and valence (v) band. The {{dashed} blue} lines are the band
  structures without EPI. The black lines are obtained by renormalizing the band according to
  the mass enhancement in Fig.~\ref{fig1}(c).
  {To facilitate the rendering, the quasiparticle spectral function is calculated
  after convoluting the self-energy in Eq.~(\ref{eq:model}) with a 5~meV Gaussian.
  The energy of the electron-phonon band replicas could be improved by using the cumulant
  expansion~\cite{Aryasetiawan1996,Guzzo2011,Lischner2013,gumh16,Verdi2017,Nery2018}{,}
  but the qualitative picture would not change.}
  }
  \label{fig3}
\end{figure}
Figure~\ref{fig3} shows the quasiparticle spectral function calculated for CH$_3$NH$_3$PbI$_3$ 
using the self-energy in Eq.~(\ref{eq:model}).
The spectral function is obtained from the self-energy via 
$ A({\bf k},\varepsilon)= -\pi^{-1}\sum_n \,{\rm Im}\,[\,\varepsilon-\varepsilon_{n{\bf k}}
-\Sigma_{n{\bf k}}(\varepsilon)]^{-1}$, and represents the many-body ${\bf k}$-resolved density
of states, i.e.{,} the many-body band structure of the system.
In the absence of EPI the bands are sharp and parabolic near the 
band edges. When EPI is taken into account, we find {strongly renormalized 
but weakly damped carriers for energies below $\sim$22~meV, 
and weakly renormalized but heavily damped 
carriers above this threshold}. 
In addition we find replica bands that are reminiscent of the polaronic satellites
recently observed in oxide perovskites and other transition metals oxides
\cite{Moser2013, Chen2015a, Cancellieri2016, Wang2016, Verdi2017, Nery2018}.
{The features} in the valence bands
should be observable via high-resolution angle-resolved 
photoelectron spectroscopy experiments {on single-crystal CH$_3$NH$_3$PbI$_3$ samples}. 

We note that, even though our calculations of the EPI matrix elements include 
spin-orbit coupling, we do not observe a dynamic Rashba-Dresselhaus {spin splitting} at the band edges.
The absence of {spin splitting} is {consistent with
the inversion symmetry of the crystal}.

%\section{Discussion}
Since polar modes dominate the 
EPI in CH{$_3$}NH$_3$PbI$_3$, we can discuss its polaronic properties starting from the 
mass enhancement parameter.
The calculated parameter $\lambda = 0.28$ is relatively small{; therefore,} we can determine the polaron 
coupling strength $\alpha$ using the {weak-coupling} expansion of Feynman's polaron mass 
\cite{Devreese2009}: $1+\alpha/6+0.025\alpha^2 
= 1+\lambda$.  We obtain $\alpha = 1.4$, which falls in the weak-coupling regime, as anticipated.
This polaron coupling strength can be used to determine the polaron binding energy $E_{\rm p}$ and its
radius $r_{\rm p}$. From Feynman's model we have $E_{\rm p} = \alpha (1 + 0.0123\,\alpha) \hbar\omega$~\cite{Feynman1955}
and $r_{\rm p} = (3.4\,\hbar/m^* \,\alpha\, \omega)^{1/2}$~\cite{Schultz1959}; using $\hbar\omega = $13-22~meV from the above analysis and $m^*=0.22$~\cite{fvg15} 
we find $E_{\rm p} = {19-31}$~meV and $r_{\rm p} = {62-81}$~\AA.
%
% r_p (A) = 161.1 / sqrt( alpha * (m*/me) *omega [meV] )
%
These results indicate that the electron-phonon coupling in CH$_3$NH$_3$PbI$_3$ leads to the formation
of large polarons {extending} over more than 20 PbI$_6$ octahedra. The binding energy of these
quasiparticles is comparable to the vibrational energy of the polar modes at room temperature{;
%
% the vibrational energy of each phonon mode is n*hbar omega, n = 1/[exp(en/kt)-1]
% for w = 13 meV and kt = 26 meV we have n = 1.54 and n*hbar omega = 20 meV 
%
therefore,} we do not expect any localization or self-trapping under standard operating conditions.
On the other hand, we note that the polaron binding energy is similar to the exciton binding energy
$E_{x} =20\pm 2$~meV in this compound~\cite{Davies2018}, therefore {polarons may}
play a role in the excitonic physics of halide perovskites.

In conclusion, we presented the first complete many-body investigation of electron-phonon physics in 
CH$_3$NH$_3$PbI$_3$. We found evidence for a novel
{multiphonon} Fr\"ohlich coupling, and used this to rationalize 
a number of experimental observations. We established that the EPI leads to ultrafast carrier 
relaxation near the band edges, and a moderate renormalization of the effective masses. Our analysis 
indicates that this system is in the weak polaronic regime. 
More generally, the {multiphonon} Fr\"ohlich model 
that we developed to examine our {\it ab initio} data can be used to investigate electron-phonon 
physics in the broader family of halide perovskites, and to establish design rules 
for engineering carrier dynamics in this {promising} class of semiconductors.

%\section{Acknowledgments}

\begin{acknowledgments}
The research leading to these results has received funding from the Leverhulme Trust (Grant {No.} RL-2012-001), 
the UK Engineering and Physical Sciences Research Council ({Grant} No. EP/M020517/1), and the {Graphene Flagship (Horizon 2020 Grant No. 785219 - GrapheneCore2)}. The authors acknowledge the use of the University of Oxford Advanced 
Research Computing (ARC) facility, the ARCHER UK National 
Supercomputing Service under the \emph{T-Dops} project, and the Cambridge Service for Data Driven Discovery ({Grant No.} EP/P020259/1). We acknowledge 
PRACE for awarding us access to Cartesius at SURFsara, Netherlands; Abel at UiO, Norway; and MareNostrum at BSC-CNS, Spain. Structural models were rendered using 
{\scshape vesta}~\cite{vesta}.
\end{acknowledgments}

%\section{Author contributions}
%M.S. and S.P. performed the computational research, M.S. analyzed the results and developed
%the analytical model, S.P. improved the EPW code, and F.G. designed the research and led
%the project.
%All authors participated in the preparation of the manuscript.
%
%\section{Additional information}
%{\bf Supplementary Information} accompanies this paper at [link]
%
%{\bf Competing financial interests:} The authors declare no competing financial interest.

%\bibliographystyle{nature}
\nocite{perez15,QE2017,wannier90,epw,ca80,pz81,hama13,sg15,Lejaeghere2016,Setten2018,gcl07,
rycr09,fvg15,vg15,Sjakste2015,Marronnier2017,kara15,wrig16,Perez2017,Seeger2004,
Grimvall1981,Galkowski2016,froe54}
\nocite{Mahan2000,Devreese2009}

\bibliography{biblio}

\begin{thebibliography}{82}%
\makeatletter
\providecommand \@ifxundefined [1]{%
 \@ifx{#1\undefined}
}%
\providecommand \@ifnum [1]{%
 \ifnum #1\expandafter \@firstoftwo
 \else \expandafter \@secondoftwo
 \fi
}%
\providecommand \@ifx [1]{%
 \ifx #1\expandafter \@firstoftwo
 \else \expandafter \@secondoftwo
 \fi
}%
\providecommand \natexlab [1]{#1}%
\providecommand \enquote  [1]{``#1''}%
\providecommand \bibnamefont  [1]{#1}%
\providecommand \bibfnamefont [1]{#1}%
\providecommand \citenamefont [1]{#1}%
\providecommand \href@noop [0]{\@secondoftwo}%
\providecommand \href [0]{\begingroup \@sanitize@url \@href}%
\providecommand \@href[1]{\@@startlink{#1}\@@href}%
\providecommand \@@href[1]{\endgroup#1\@@endlink}%
\providecommand \@sanitize@url [0]{\catcode `\\12\catcode `\$12\catcode
  `\&12\catcode `\#12\catcode `\^12\catcode `\_12\catcode `\%12\relax}%
\providecommand \@@startlink[1]{}%
\providecommand \@@endlink[0]{}%
\providecommand \url  [0]{\begingroup\@sanitize@url \@url }%
\providecommand \@url [1]{\endgroup\@href {#1}{\urlprefix }}%
\providecommand \urlprefix  [0]{URL }%
\providecommand \Eprint [0]{\href }%
\providecommand \doibase [0]{http://dx.doi.org/}%
\providecommand \selectlanguage [0]{\@gobble}%
\providecommand \bibinfo  [0]{\@secondoftwo}%
\providecommand \bibfield  [0]{\@secondoftwo}%
\providecommand \translation [1]{[#1]}%
\providecommand \BibitemOpen [0]{}%
\providecommand \bibitemStop [0]{}%
\providecommand \bibitemNoStop [0]{.\EOS\space}%
\providecommand \EOS [0]{\spacefactor3000\relax}%
\providecommand \BibitemShut  [1]{\csname bibitem#1\endcsname}%
\let\auto@bib@innerbib\@empty
%</preamble>
\bibitem [{\citenamefont {Lee}\ \emph {et~al.}(2012)\citenamefont {Lee},
  \citenamefont {Teuscher}, \citenamefont {Miyasaka}, \citenamefont
  {Murakami},\ and\ \citenamefont {Snaith}}]{lee12}%
  \BibitemOpen
  \bibfield  {author} {\bibinfo {author} {\bibfnamefont {M.~M.}\ \bibnamefont
  {Lee}}, \bibinfo {author} {\bibfnamefont {J.}~\bibnamefont {Teuscher}},
  \bibinfo {author} {\bibfnamefont {T.}~\bibnamefont {Miyasaka}}, \bibinfo
  {author} {\bibfnamefont {T.~N.}\ \bibnamefont {Murakami}}, \ and\ \bibinfo
  {author} {\bibfnamefont {H.~J.}\ \bibnamefont {Snaith}},\ }\href {\doibase
  10.1126/science.1228604} {\bibfield  {journal} {\bibinfo  {journal}
  {Science}\ }\textbf {\bibinfo {volume} {338}},\ \bibinfo {pages} {643}
  (\bibinfo {year} {2012})}\BibitemShut {NoStop}%
\bibitem [{\citenamefont {Kim}\ \emph {et~al.}(2012)\citenamefont {Kim},
  \citenamefont {Lee}, \citenamefont {Im}, \citenamefont {Lee}, \citenamefont
  {Moehl}, \citenamefont {Marchioro}, \citenamefont {Moon}, \citenamefont
  {Humphry-Baker}, \citenamefont {Yum}, \citenamefont {Moser}, \citenamefont
  {Gr\"atzel},\ and\ \citenamefont {Park}}]{kim12}%
  \BibitemOpen
  \bibfield  {author} {\bibinfo {author} {\bibfnamefont {H.-S.}\ \bibnamefont
  {Kim}}, \bibinfo {author} {\bibfnamefont {C.-R.}\ \bibnamefont {Lee}},
  \bibinfo {author} {\bibfnamefont {J.-H.}\ \bibnamefont {Im}}, \bibinfo
  {author} {\bibfnamefont {K.-B.}\ \bibnamefont {Lee}}, \bibinfo {author}
  {\bibfnamefont {T.}~\bibnamefont {Moehl}}, \bibinfo {author} {\bibfnamefont
  {A.}~\bibnamefont {Marchioro}}, \bibinfo {author} {\bibfnamefont {S.-J.}\
  \bibnamefont {Moon}}, \bibinfo {author} {\bibfnamefont {R.}~\bibnamefont
  {Humphry-Baker}}, \bibinfo {author} {\bibfnamefont {J.-H.}\ \bibnamefont
  {Yum}}, \bibinfo {author} {\bibfnamefont {J.~E.}\ \bibnamefont {Moser}},
  \bibinfo {author} {\bibfnamefont {M.}~\bibnamefont {Gr\"atzel}}, \ and\
  \bibinfo {author} {\bibfnamefont {N.-G.}\ \bibnamefont {Park}},\ }\href
  {http://dx.doi.org/10.1038/srep00591} {\bibfield  {journal} {\bibinfo
  {journal} {Sci. Rep.}\ }\textbf {\bibinfo {volume} {2}},\ \bibinfo {pages}
  {591} (\bibinfo {year} {2012})}\BibitemShut {NoStop}%
\bibitem [{\citenamefont {Yablonovitch}\ \emph {et~al.}(2012)\citenamefont
  {Yablonovitch}, \citenamefont {Miller},\ and\ \citenamefont
  {Kurtz}}]{Yablonovitch2012}%
  \BibitemOpen
  \bibfield  {author} {\bibinfo {author} {\bibfnamefont {E.}~\bibnamefont
  {Yablonovitch}}, \bibinfo {author} {\bibfnamefont {O.~D.}\ \bibnamefont
  {Miller}}, \ and\ \bibinfo {author} {\bibfnamefont {S.~R.}\ \bibnamefont
  {Kurtz}},\ }in\ \href {\doibase 10.1109/PVSC.2012.6317891} {\emph {\bibinfo
  {booktitle} {Proceedings of the 38th IEEE Photovoltaic Specialists
  Conference}}}\ (\bibinfo {year} {2012})\ pp.\ \bibinfo {pages}
  {001556--001559}\BibitemShut {NoStop}%
\bibitem [{\citenamefont {Green}\ \emph {et~al.}(2018)\citenamefont {Green},
  \citenamefont {Hishikawa}, \citenamefont {Dunlop}, \citenamefont {Levi},
  \citenamefont {Hohl-Ebinger},\ and\ \citenamefont {Ho-Baillie}}]{Green2018}%
  \BibitemOpen
  \bibfield  {author} {\bibinfo {author} {\bibfnamefont {M.~A.}\ \bibnamefont
  {Green}}, \bibinfo {author} {\bibfnamefont {Y.}~\bibnamefont {Hishikawa}},
  \bibinfo {author} {\bibfnamefont {E.~D.}\ \bibnamefont {Dunlop}}, \bibinfo
  {author} {\bibfnamefont {D.~H.}\ \bibnamefont {Levi}}, \bibinfo {author}
  {\bibfnamefont {J.}~\bibnamefont {Hohl-Ebinger}}, \ and\ \bibinfo {author}
  {\bibfnamefont {A.~W.}\ \bibnamefont {Ho-Baillie}},\ }\href {\doibase
  10.1002/pip.2978} {\bibfield  {journal} {\bibinfo  {journal} {Prog.
  Photovoltaics Res. Appl.}\ }\textbf {\bibinfo {volume} {26}},\ \bibinfo
  {pages} {3} (\bibinfo {year} {2018})}\BibitemShut {NoStop}%
\bibitem [{\citenamefont {Green}\ \emph {et~al.}(2014)\citenamefont {Green},
  \citenamefont {Ho-Baillie},\ and\ \citenamefont {Snaith}}]{ghs14}%
  \BibitemOpen
  \bibfield  {author} {\bibinfo {author} {\bibfnamefont {M.~A.}\ \bibnamefont
  {Green}}, \bibinfo {author} {\bibfnamefont {A.}~\bibnamefont {Ho-Baillie}}, \
  and\ \bibinfo {author} {\bibfnamefont {H.~J.}\ \bibnamefont {Snaith}},\
  }\href {http://dx.doi.org/10.1038/nphoton.2014.134} {\bibfield  {journal}
  {\bibinfo  {journal} {Nat. Photonics}\ }\textbf {\bibinfo {volume} {8}},\
  \bibinfo {pages} {506} (\bibinfo {year} {2014})}\BibitemShut {NoStop}%
\bibitem [{\citenamefont {Johnston}\ and\ \citenamefont
  {Herz}(2016)}]{Johnston2016}%
  \BibitemOpen
  \bibfield  {author} {\bibinfo {author} {\bibfnamefont {M.~B.}\ \bibnamefont
  {Johnston}}\ and\ \bibinfo {author} {\bibfnamefont {L.~M.}\ \bibnamefont
  {Herz}},\ }\href {\doibase 10.1021/acs.accounts.5b00411} {\bibfield
  {journal} {\bibinfo  {journal} {Acc. Chem. Res.}\ }\textbf {\bibinfo {volume}
  {49}},\ \bibinfo {pages} {146} (\bibinfo {year} {2016})}\BibitemShut
  {NoStop}%
\bibitem [{\citenamefont {Herz}(2017)}]{Herz2017}%
  \BibitemOpen
  \bibfield  {author} {\bibinfo {author} {\bibfnamefont {L.~M.}\ \bibnamefont
  {Herz}},\ }\href {\doibase 10.1021/acsenergylett.7b00276} {\bibfield
  {journal} {\bibinfo  {journal} {ACS Energy Lett.}\ }\textbf {\bibinfo
  {volume} {2}},\ \bibinfo {pages} {1539} (\bibinfo {year} {2017})}\BibitemShut
  {NoStop}%
\bibitem [{\citenamefont {Wehrenfennig}\ \emph
  {et~al.}(2014{\natexlab{a}})\citenamefont {Wehrenfennig}, \citenamefont
  {Eperon}, \citenamefont {Johnston}, \citenamefont {Snaith},\ and\
  \citenamefont {Herz}}]{wehr14}%
  \BibitemOpen
  \bibfield  {author} {\bibinfo {author} {\bibfnamefont {C.}~\bibnamefont
  {Wehrenfennig}}, \bibinfo {author} {\bibfnamefont {G.~E.}\ \bibnamefont
  {Eperon}}, \bibinfo {author} {\bibfnamefont {M.~B.}\ \bibnamefont
  {Johnston}}, \bibinfo {author} {\bibfnamefont {H.~J.}\ \bibnamefont
  {Snaith}}, \ and\ \bibinfo {author} {\bibfnamefont {L.~M.}\ \bibnamefont
  {Herz}},\ }\href {\doibase 10.1002/adma.201305172} {\bibfield  {journal}
  {\bibinfo  {journal} {Adv. Mater.}\ }\textbf {\bibinfo {volume} {26}},\
  \bibinfo {pages} {1584} (\bibinfo {year} {2014}{\natexlab{a}})}\BibitemShut
  {NoStop}%
\bibitem [{\citenamefont {Wehrenfennig}\ \emph
  {et~al.}(2014{\natexlab{b}})\citenamefont {Wehrenfennig}, \citenamefont
  {Liu}, \citenamefont {Snaith}, \citenamefont {Johnston},\ and\ \citenamefont
  {Herz}}]{Wehrenfennig2014}%
  \BibitemOpen
  \bibfield  {author} {\bibinfo {author} {\bibfnamefont {C.}~\bibnamefont
  {Wehrenfennig}}, \bibinfo {author} {\bibfnamefont {M.}~\bibnamefont {Liu}},
  \bibinfo {author} {\bibfnamefont {H.~J.}\ \bibnamefont {Snaith}}, \bibinfo
  {author} {\bibfnamefont {M.~B.}\ \bibnamefont {Johnston}}, \ and\ \bibinfo
  {author} {\bibfnamefont {L.~M.}\ \bibnamefont {Herz}},\ }\href {\doibase
  10.1021/jz500434p} {\bibfield  {journal} {\bibinfo  {journal} {J. Phys. Chem.
  Lett.}\ }\textbf {\bibinfo {volume} {5}},\ \bibinfo {pages} {1300} (\bibinfo
  {year} {2014}{\natexlab{b}})}\BibitemShut {NoStop}%
\bibitem [{\citenamefont {Karakus}\ \emph {et~al.}(2015)\citenamefont
  {Karakus}, \citenamefont {Jensen}, \citenamefont {D'Angelo}, \citenamefont
  {Turchinovich}, \citenamefont {Bonn},\ and\ \citenamefont
  {C\'anovas}}]{kara15}%
  \BibitemOpen
  \bibfield  {author} {\bibinfo {author} {\bibfnamefont {M.}~\bibnamefont
  {Karakus}}, \bibinfo {author} {\bibfnamefont {S.~A.}\ \bibnamefont {Jensen}},
  \bibinfo {author} {\bibfnamefont {F.}~\bibnamefont {D'Angelo}}, \bibinfo
  {author} {\bibfnamefont {D.}~\bibnamefont {Turchinovich}}, \bibinfo {author}
  {\bibfnamefont {M.}~\bibnamefont {Bonn}}, \ and\ \bibinfo {author}
  {\bibfnamefont {E.}~\bibnamefont {C\'anovas}},\ }\href {\doibase
  10.1021/acs.jpclett.5b02485} {\bibfield  {journal} {\bibinfo  {journal} {J.
  Phys. Chem. Lett.}\ }\textbf {\bibinfo {volume} {6}},\ \bibinfo {pages}
  {4991} (\bibinfo {year} {2015})}\BibitemShut {NoStop}%
\bibitem [{\citenamefont {Milot}\ \emph {et~al.}(2015)\citenamefont {Milot},
  \citenamefont {Eperon}, \citenamefont {Snaith}, \citenamefont {Johnston},\
  and\ \citenamefont {Herz}}]{milot15}%
  \BibitemOpen
  \bibfield  {author} {\bibinfo {author} {\bibfnamefont {R.~L.}\ \bibnamefont
  {Milot}}, \bibinfo {author} {\bibfnamefont {G.~E.}\ \bibnamefont {Eperon}},
  \bibinfo {author} {\bibfnamefont {H.~J.}\ \bibnamefont {Snaith}}, \bibinfo
  {author} {\bibfnamefont {M.~B.}\ \bibnamefont {Johnston}}, \ and\ \bibinfo
  {author} {\bibfnamefont {L.~M.}\ \bibnamefont {Herz}},\ }\href {\doibase
  10.1002/adfm.201502340} {\bibfield  {journal} {\bibinfo  {journal} {Adv.
  Funct. Mater.}\ }\textbf {\bibinfo {volume} {25}},\ \bibinfo {pages} {6218}
  (\bibinfo {year} {2015})}\BibitemShut {NoStop}%
\bibitem [{\citenamefont {Herz}(2016)}]{herz16}%
  \BibitemOpen
  \bibfield  {author} {\bibinfo {author} {\bibfnamefont {L.~M.}\ \bibnamefont
  {Herz}},\ }\href {\doibase 10.1146/annurev-physchem-040215-112222} {\bibfield
   {journal} {\bibinfo  {journal} {Annu. Rev. Phys. Chem.}\ }\textbf {\bibinfo
  {volume} {67}},\ \bibinfo {pages} {65} (\bibinfo {year} {2016})}\BibitemShut
  {NoStop}%
\bibitem [{\citenamefont {Filippetti}\ \emph {et~al.}(2016)\citenamefont
  {Filippetti}, \citenamefont {Mattoni}, \citenamefont {Caddeo}, \citenamefont
  {Saba},\ and\ \citenamefont {Delugas}}]{Filippetti2016}%
  \BibitemOpen
  \bibfield  {author} {\bibinfo {author} {\bibfnamefont {A.}~\bibnamefont
  {Filippetti}}, \bibinfo {author} {\bibfnamefont {A.}~\bibnamefont {Mattoni}},
  \bibinfo {author} {\bibfnamefont {C.}~\bibnamefont {Caddeo}}, \bibinfo
  {author} {\bibfnamefont {M.~I.}\ \bibnamefont {Saba}}, \ and\ \bibinfo
  {author} {\bibfnamefont {P.}~\bibnamefont {Delugas}},\ }\href {\doibase
  10.1039/C6CP01402J} {\bibfield  {journal} {\bibinfo  {journal} {Phys. Chem.
  Chem. Phys.}\ }\textbf {\bibinfo {volume} {18}},\ \bibinfo {pages} {15352}
  (\bibinfo {year} {2016})}\BibitemShut {NoStop}%
\bibitem [{\citenamefont {Wu}\ \emph {et~al.}(2014)\citenamefont {Wu},
  \citenamefont {Bera}, \citenamefont {Ma}, \citenamefont {Du}, \citenamefont
  {Yang}, \citenamefont {Li},\ and\ \citenamefont {Wu}}]{Wu2014}%
  \BibitemOpen
  \bibfield  {author} {\bibinfo {author} {\bibfnamefont {K.}~\bibnamefont
  {Wu}}, \bibinfo {author} {\bibfnamefont {A.}~\bibnamefont {Bera}}, \bibinfo
  {author} {\bibfnamefont {C.}~\bibnamefont {Ma}}, \bibinfo {author}
  {\bibfnamefont {Y.}~\bibnamefont {Du}}, \bibinfo {author} {\bibfnamefont
  {Y.}~\bibnamefont {Yang}}, \bibinfo {author} {\bibfnamefont {L.}~\bibnamefont
  {Li}}, \ and\ \bibinfo {author} {\bibfnamefont {T.}~\bibnamefont {Wu}},\
  }\href {\doibase 10.1039/C4CP03573A} {\bibfield  {journal} {\bibinfo
  {journal} {Phys. Chem. Chem. Phys.}\ }\textbf {\bibinfo {volume} {16}},\
  \bibinfo {pages} {22476} (\bibinfo {year} {2014})}\BibitemShut {NoStop}%
\bibitem [{\citenamefont {Dar}\ \emph {et~al.}(2016)\citenamefont {Dar},
  \citenamefont {Jacopin}, \citenamefont {Meloni}, \citenamefont {Mattoni},
  \citenamefont {Arora}, \citenamefont {Boziki}, \citenamefont {Zakeeruddin},
  \citenamefont {Rothlisberger},\ and\ \citenamefont {Gr{\"a}tzel}}]{dar16}%
  \BibitemOpen
  \bibfield  {author} {\bibinfo {author} {\bibfnamefont {M.~I.}\ \bibnamefont
  {Dar}}, \bibinfo {author} {\bibfnamefont {G.}~\bibnamefont {Jacopin}},
  \bibinfo {author} {\bibfnamefont {S.}~\bibnamefont {Meloni}}, \bibinfo
  {author} {\bibfnamefont {A.}~\bibnamefont {Mattoni}}, \bibinfo {author}
  {\bibfnamefont {N.}~\bibnamefont {Arora}}, \bibinfo {author} {\bibfnamefont
  {A.}~\bibnamefont {Boziki}}, \bibinfo {author} {\bibfnamefont {S.~M.}\
  \bibnamefont {Zakeeruddin}}, \bibinfo {author} {\bibfnamefont
  {U.}~\bibnamefont {Rothlisberger}}, \ and\ \bibinfo {author} {\bibfnamefont
  {M.}~\bibnamefont {Gr{\"a}tzel}},\ }\href {\doibase 10.1126/sciadv.1601156}
  {\bibfield  {journal} {\bibinfo  {journal} {Sci. Adv.}\ }\textbf {\bibinfo
  {volume} {2}},\ \bibinfo {pages} {e1601156} (\bibinfo {year}
  {2016})}\BibitemShut {NoStop}%
\bibitem [{\citenamefont {Zheng}\ \emph {et~al.}(2015)\citenamefont {Zheng},
  \citenamefont {Tan}, \citenamefont {Liu},\ and\ \citenamefont
  {Rappe}}]{ztlr15}%
  \BibitemOpen
  \bibfield  {author} {\bibinfo {author} {\bibfnamefont {F.}~\bibnamefont
  {Zheng}}, \bibinfo {author} {\bibfnamefont {L.~Z.}\ \bibnamefont {Tan}},
  \bibinfo {author} {\bibfnamefont {S.}~\bibnamefont {Liu}}, \ and\ \bibinfo
  {author} {\bibfnamefont {A.~M.}\ \bibnamefont {Rappe}},\ }\href {\doibase
  10.1021/acs.nanolett.5b01854} {\bibfield  {journal} {\bibinfo  {journal}
  {Nano Lett.}\ }\textbf {\bibinfo {volume} {15}},\ \bibinfo {pages} {7794}
  (\bibinfo {year} {2015})}\BibitemShut {NoStop}%
\bibitem [{\citenamefont {Kepenekian}\ \emph {et~al.}(2015)\citenamefont
  {Kepenekian}, \citenamefont {Robles}, \citenamefont {Katan}, \citenamefont
  {Sapori}, \citenamefont {Pedesseau},\ and\ \citenamefont {Even}}]{kepe15}%
  \BibitemOpen
  \bibfield  {author} {\bibinfo {author} {\bibfnamefont {M.}~\bibnamefont
  {Kepenekian}}, \bibinfo {author} {\bibfnamefont {R.}~\bibnamefont {Robles}},
  \bibinfo {author} {\bibfnamefont {C.}~\bibnamefont {Katan}}, \bibinfo
  {author} {\bibfnamefont {D.}~\bibnamefont {Sapori}}, \bibinfo {author}
  {\bibfnamefont {L.}~\bibnamefont {Pedesseau}}, \ and\ \bibinfo {author}
  {\bibfnamefont {J.}~\bibnamefont {Even}},\ }\href {\doibase
  10.1021/acsnano.5b04409} {\bibfield  {journal} {\bibinfo  {journal} {ACS
  Nano}\ }\textbf {\bibinfo {volume} {9}},\ \bibinfo {pages} {11557} (\bibinfo
  {year} {2015})}\BibitemShut {NoStop}%
\bibitem [{\citenamefont {Etienne}\ \emph {et~al.}(2016)\citenamefont
  {Etienne}, \citenamefont {Mosconi},\ and\ \citenamefont
  {De~Angelis}}]{emd16}%
  \BibitemOpen
  \bibfield  {author} {\bibinfo {author} {\bibfnamefont {T.}~\bibnamefont
  {Etienne}}, \bibinfo {author} {\bibfnamefont {E.}~\bibnamefont {Mosconi}}, \
  and\ \bibinfo {author} {\bibfnamefont {F.}~\bibnamefont {De~Angelis}},\
  }\href {\doibase 10.1021/acs.jpclett.6b00564} {\bibfield  {journal} {\bibinfo
   {journal} {J. Phys. Chem. Lett.}\ }\textbf {\bibinfo {volume} {7}},\
  \bibinfo {pages} {1638} (\bibinfo {year} {2016})}\BibitemShut {NoStop}%
\bibitem [{\citenamefont {Price}\ \emph {et~al.}(2015)\citenamefont {Price},
  \citenamefont {Butkus}, \citenamefont {Jellicoe}, \citenamefont {Sadhanala},
  \citenamefont {Briane}, \citenamefont {Halpert}, \citenamefont {Broch},
  \citenamefont {Hodgkiss}, \citenamefont {Friend},\ and\ \citenamefont
  {Deschler}}]{Price2015}%
  \BibitemOpen
  \bibfield  {author} {\bibinfo {author} {\bibfnamefont {M.~B.}\ \bibnamefont
  {Price}}, \bibinfo {author} {\bibfnamefont {J.}~\bibnamefont {Butkus}},
  \bibinfo {author} {\bibfnamefont {T.~C.}\ \bibnamefont {Jellicoe}}, \bibinfo
  {author} {\bibfnamefont {A.}~\bibnamefont {Sadhanala}}, \bibinfo {author}
  {\bibfnamefont {A.}~\bibnamefont {Briane}}, \bibinfo {author} {\bibfnamefont
  {J.~E.}\ \bibnamefont {Halpert}}, \bibinfo {author} {\bibfnamefont
  {K.}~\bibnamefont {Broch}}, \bibinfo {author} {\bibfnamefont {J.~M.}\
  \bibnamefont {Hodgkiss}}, \bibinfo {author} {\bibfnamefont {R.~H.}\
  \bibnamefont {Friend}}, \ and\ \bibinfo {author} {\bibfnamefont
  {F.}~\bibnamefont {Deschler}},\ }\href {http://dx.doi.org/10.1038/ncomms9420}
  {\bibfield  {journal} {\bibinfo  {journal} {Nat. Commun.}\ }\textbf {\bibinfo
  {volume} {6}},\ \bibinfo {pages} {8420} (\bibinfo {year} {2015})}\BibitemShut
  {NoStop}%
\bibitem [{\citenamefont {Yang}\ \emph {et~al.}(2016)\citenamefont {Yang},
  \citenamefont {Ostrowski}, \citenamefont {France}, \citenamefont {Zhu},
  \citenamefont {van~de Lagemaat}, \citenamefont {Luther},\ and\ \citenamefont
  {Beard}}]{Yang2016}%
  \BibitemOpen
  \bibfield  {author} {\bibinfo {author} {\bibfnamefont {Y.}~\bibnamefont
  {Yang}}, \bibinfo {author} {\bibfnamefont {D.~P.}\ \bibnamefont {Ostrowski}},
  \bibinfo {author} {\bibfnamefont {R.~M.}\ \bibnamefont {France}}, \bibinfo
  {author} {\bibfnamefont {K.}~\bibnamefont {Zhu}}, \bibinfo {author}
  {\bibfnamefont {J.}~\bibnamefont {van~de Lagemaat}}, \bibinfo {author}
  {\bibfnamefont {J.~M.}\ \bibnamefont {Luther}}, \ and\ \bibinfo {author}
  {\bibfnamefont {M.~C.}\ \bibnamefont {Beard}},\ }\href
  {http://dx.doi.org/10.1038/nphoton.2015.213} {\bibfield  {journal} {\bibinfo
  {journal} {Nat. Photonics}\ }\textbf {\bibinfo {volume} {10}},\ \bibinfo
  {pages} {53} (\bibinfo {year} {2016})}\BibitemShut {NoStop}%
\bibitem [{\citenamefont {Yang}\ \emph {et~al.}(2017)\citenamefont {Yang},
  \citenamefont {Wen}, \citenamefont {Xia}, \citenamefont {Sheng},
  \citenamefont {Ma}, \citenamefont {Kim}, \citenamefont {Tapping},
  \citenamefont {Harada}, \citenamefont {Kee}, \citenamefont {Huang},
  \citenamefont {Cheng}, \citenamefont {Green}, \citenamefont {Ho-Baillie},
  \citenamefont {Huang}, \citenamefont {Shrestha}, \citenamefont {Patterson},\
  and\ \citenamefont {Conibeer}}]{Yang2017b}%
  \BibitemOpen
  \bibfield  {author} {\bibinfo {author} {\bibfnamefont {J.}~\bibnamefont
  {Yang}}, \bibinfo {author} {\bibfnamefont {X.}~\bibnamefont {Wen}}, \bibinfo
  {author} {\bibfnamefont {H.}~\bibnamefont {Xia}}, \bibinfo {author}
  {\bibfnamefont {R.}~\bibnamefont {Sheng}}, \bibinfo {author} {\bibfnamefont
  {Q.}~\bibnamefont {Ma}}, \bibinfo {author} {\bibfnamefont {J.}~\bibnamefont
  {Kim}}, \bibinfo {author} {\bibfnamefont {P.}~\bibnamefont {Tapping}},
  \bibinfo {author} {\bibfnamefont {T.}~\bibnamefont {Harada}}, \bibinfo
  {author} {\bibfnamefont {T.~W.}\ \bibnamefont {Kee}}, \bibinfo {author}
  {\bibfnamefont {F.}~\bibnamefont {Huang}}, \bibinfo {author} {\bibfnamefont
  {Y.-B.}\ \bibnamefont {Cheng}}, \bibinfo {author} {\bibfnamefont
  {M.}~\bibnamefont {Green}}, \bibinfo {author} {\bibfnamefont
  {A.}~\bibnamefont {Ho-Baillie}}, \bibinfo {author} {\bibfnamefont
  {S.}~\bibnamefont {Huang}}, \bibinfo {author} {\bibfnamefont
  {S.}~\bibnamefont {Shrestha}}, \bibinfo {author} {\bibfnamefont
  {R.}~\bibnamefont {Patterson}}, \ and\ \bibinfo {author} {\bibfnamefont
  {G.}~\bibnamefont {Conibeer}},\ }\href
  {http://dx.doi.org/10.1038/ncomms14120} {\bibfield  {journal} {\bibinfo
  {journal} {Nat. Commun.}\ }\textbf {\bibinfo {volume} {8}},\ \bibinfo {pages}
  {14120} (\bibinfo {year} {2017})}\BibitemShut {NoStop}%
\bibitem [{\citenamefont {Wright}\ \emph {et~al.}(2016)\citenamefont {Wright},
  \citenamefont {Verdi}, \citenamefont {Milot}, \citenamefont {Eperon},
  \citenamefont {P\'erez-Osorio}, \citenamefont {Snaith}, \citenamefont
  {Giustino}, \citenamefont {Johnston},\ and\ \citenamefont {Herz}}]{wrig16}%
  \BibitemOpen
  \bibfield  {author} {\bibinfo {author} {\bibfnamefont {A.~D.}\ \bibnamefont
  {Wright}}, \bibinfo {author} {\bibfnamefont {C.}~\bibnamefont {Verdi}},
  \bibinfo {author} {\bibfnamefont {R.~L.}\ \bibnamefont {Milot}}, \bibinfo
  {author} {\bibfnamefont {G.~E.}\ \bibnamefont {Eperon}}, \bibinfo {author}
  {\bibfnamefont {M.~A.}\ \bibnamefont {P\'erez-Osorio}}, \bibinfo {author}
  {\bibfnamefont {H.~J.}\ \bibnamefont {Snaith}}, \bibinfo {author}
  {\bibfnamefont {F.}~\bibnamefont {Giustino}}, \bibinfo {author}
  {\bibfnamefont {M.~B.}\ \bibnamefont {Johnston}}, \ and\ \bibinfo {author}
  {\bibfnamefont {L.~M.}\ \bibnamefont {Herz}},\ }\href {\doibase
  doi:10.1038/ncomms11755} {\bibfield  {journal} {\bibinfo  {journal} {Nat.
  Commun.}\ }\textbf {\bibinfo {volume} {7}},\ \bibinfo {pages} {11755}
  (\bibinfo {year} {2016})}\BibitemShut {NoStop}%
\bibitem [{\citenamefont {P\'erez-Osorio}\ \emph {et~al.}(2015)\citenamefont
  {P\'erez-Osorio}, \citenamefont {Milot}, \citenamefont {Filip}, \citenamefont
  {Patel}, \citenamefont {Herz}, \citenamefont {Johnston},\ and\ \citenamefont
  {Giustino}}]{perez15}%
  \BibitemOpen
  \bibfield  {author} {\bibinfo {author} {\bibfnamefont {M.~A.}\ \bibnamefont
  {P\'erez-Osorio}}, \bibinfo {author} {\bibfnamefont {R.~L.}\ \bibnamefont
  {Milot}}, \bibinfo {author} {\bibfnamefont {M.~R.}\ \bibnamefont {Filip}},
  \bibinfo {author} {\bibfnamefont {J.~B.}\ \bibnamefont {Patel}}, \bibinfo
  {author} {\bibfnamefont {L.~M.}\ \bibnamefont {Herz}}, \bibinfo {author}
  {\bibfnamefont {M.~B.}\ \bibnamefont {Johnston}}, \ and\ \bibinfo {author}
  {\bibfnamefont {F.}~\bibnamefont {Giustino}},\ }\href {\doibase
  10.1021/acs.jpcc.5b07432} {\bibfield  {journal} {\bibinfo  {journal} {J.
  Phys. Chem. C}\ }\textbf {\bibinfo {volume} {119}},\ \bibinfo {pages} {25703}
  (\bibinfo {year} {2015})}\BibitemShut {NoStop}%
\bibitem [{\citenamefont {Even}\ \emph {et~al.}(2016)\citenamefont {Even},
  \citenamefont {Paofai}, \citenamefont {Bourges}, \citenamefont {Letoublon},
  \citenamefont {Cordier}, \citenamefont {Durand},\ and\ \citenamefont
  {Katan}}]{Even2016}%
  \BibitemOpen
  \bibfield  {author} {\bibinfo {author} {\bibfnamefont {J.}~\bibnamefont
  {Even}}, \bibinfo {author} {\bibfnamefont {S.}~\bibnamefont {Paofai}},
  \bibinfo {author} {\bibfnamefont {P.}~\bibnamefont {Bourges}}, \bibinfo
  {author} {\bibfnamefont {A.}~\bibnamefont {Letoublon}}, \bibinfo {author}
  {\bibfnamefont {S.}~\bibnamefont {Cordier}}, \bibinfo {author} {\bibfnamefont
  {O.}~\bibnamefont {Durand}}, \ and\ \bibinfo {author} {\bibfnamefont
  {C.}~\bibnamefont {Katan}},\ }\href {\doibase 10.1117/12.2213623} {\bibfield
  {journal} {\bibinfo  {journal} {Proc. SPIE}\ }\textbf {\bibinfo {volume}
  {9743}},\ \bibinfo {pages} {97430M} (\bibinfo {year} {2016})}\BibitemShut
  {NoStop}%
\bibitem [{\citenamefont {Zhu}\ and\ \citenamefont {Podzorov}(2015)}]{Zhu2015}%
  \BibitemOpen
  \bibfield  {author} {\bibinfo {author} {\bibfnamefont {X.-Y.}\ \bibnamefont
  {Zhu}}\ and\ \bibinfo {author} {\bibfnamefont {V.}~\bibnamefont {Podzorov}},\
  }\href {\doibase 10.1021/acs.jpclett.5b02462} {\bibfield  {journal} {\bibinfo
   {journal} {J. Phys. Chem. Lett.}\ }\textbf {\bibinfo {volume} {6}},\
  \bibinfo {pages} {4758} (\bibinfo {year} {2015})}\BibitemShut {NoStop}%
\bibitem [{\citenamefont {Chen}\ \emph {et~al.}(2016)\citenamefont {Chen},
  \citenamefont {Yi}, \citenamefont {Wu}, \citenamefont {Haroldson},
  \citenamefont {Gartstein}, \citenamefont {Rodionov}, \citenamefont
  {Tikhonov}, \citenamefont {Zakhidov}, \citenamefont {Zhu},\ and\
  \citenamefont {Podzorov}}]{Chen2016}%
  \BibitemOpen
  \bibfield  {author} {\bibinfo {author} {\bibfnamefont {Y.}~\bibnamefont
  {Chen}}, \bibinfo {author} {\bibfnamefont {H.~T.}\ \bibnamefont {Yi}},
  \bibinfo {author} {\bibfnamefont {X.}~\bibnamefont {Wu}}, \bibinfo {author}
  {\bibfnamefont {R.}~\bibnamefont {Haroldson}}, \bibinfo {author}
  {\bibfnamefont {Y.~N.}\ \bibnamefont {Gartstein}}, \bibinfo {author}
  {\bibfnamefont {Y.~I.}\ \bibnamefont {Rodionov}}, \bibinfo {author}
  {\bibfnamefont {K.~S.}\ \bibnamefont {Tikhonov}}, \bibinfo {author}
  {\bibfnamefont {A.}~\bibnamefont {Zakhidov}}, \bibinfo {author}
  {\bibfnamefont {X.~Y.}\ \bibnamefont {Zhu}}, \ and\ \bibinfo {author}
  {\bibfnamefont {V.}~\bibnamefont {Podzorov}},\ }\href
  {http://dx.doi.org/10.1038/ncomms12253} {\bibfield  {journal} {\bibinfo
  {journal} {Nat. Commun.}\ }\textbf {\bibinfo {volume} {7}},\ \bibinfo {pages}
  {12253} (\bibinfo {year} {2016})}\BibitemShut {NoStop}%
\bibitem [{\citenamefont {Frost}(2017)}]{Frost2017a}%
  \BibitemOpen
  \bibfield  {author} {\bibinfo {author} {\bibfnamefont {J.~M.}\ \bibnamefont
  {Frost}},\ }\href {\doibase 10.1103/PhysRevB.96.195202} {\bibfield  {journal}
  {\bibinfo  {journal} {Phys. Rev. B}\ }\textbf {\bibinfo {volume} {96}},\
  \bibinfo {pages} {195202} (\bibinfo {year} {2017})}\BibitemShut {NoStop}%
\bibitem [{\citenamefont {Zhang}\ \emph {et~al.}(2017)\citenamefont {Zhang},
  \citenamefont {Zhang}, \citenamefont {Huang}, \citenamefont {Lin},\ and\
  \citenamefont {Lu}}]{Zhang2017}%
  \BibitemOpen
  \bibfield  {author} {\bibinfo {author} {\bibfnamefont {M.}~\bibnamefont
  {Zhang}}, \bibinfo {author} {\bibfnamefont {X.}~\bibnamefont {Zhang}},
  \bibinfo {author} {\bibfnamefont {L.-Y.}\ \bibnamefont {Huang}}, \bibinfo
  {author} {\bibfnamefont {H.-Q.}\ \bibnamefont {Lin}}, \ and\ \bibinfo
  {author} {\bibfnamefont {G.}~\bibnamefont {Lu}},\ }\href {\doibase
  10.1103/PhysRevB.96.195203} {\bibfield  {journal} {\bibinfo  {journal} {Phys.
  Rev. B}\ }\textbf {\bibinfo {volume} {96}},\ \bibinfo {pages} {195203}
  (\bibinfo {year} {2017})}\BibitemShut {NoStop}%
\bibitem [{\citenamefont {Baikie}\ \emph {et~al.}(2013)\citenamefont {Baikie},
  \citenamefont {Fang}, \citenamefont {Kadro}, \citenamefont {Schreyer},
  \citenamefont {Wei}, \citenamefont {Mhaisalkar}, \citenamefont {Graetzel},\
  and\ \citenamefont {White}}]{baik13}%
  \BibitemOpen
  \bibfield  {author} {\bibinfo {author} {\bibfnamefont {T.}~\bibnamefont
  {Baikie}}, \bibinfo {author} {\bibfnamefont {Y.}~\bibnamefont {Fang}},
  \bibinfo {author} {\bibfnamefont {J.~M.}\ \bibnamefont {Kadro}}, \bibinfo
  {author} {\bibfnamefont {M.}~\bibnamefont {Schreyer}}, \bibinfo {author}
  {\bibfnamefont {F.}~\bibnamefont {Wei}}, \bibinfo {author} {\bibfnamefont
  {S.~G.}\ \bibnamefont {Mhaisalkar}}, \bibinfo {author} {\bibfnamefont
  {M.}~\bibnamefont {Graetzel}}, \ and\ \bibinfo {author} {\bibfnamefont
  {T.~J.}\ \bibnamefont {White}},\ }\href {\doibase 10.1039/C3TA10518K}
  {\bibfield  {journal} {\bibinfo  {journal} {J. Mater. Chem. A}\ }\textbf
  {\bibinfo {volume} {1}},\ \bibinfo {pages} {5628} (\bibinfo {year}
  {2013})}\BibitemShut {NoStop}%
\bibitem [{sup()}]{suppl}%
  \BibitemOpen
  \href@noop {} {}\bibinfo {note} {See Supplemental Material at
  \url{http://link.aps.org/supplemental/10.1103/PhysRevLett.121.086402} {for
  the derivation of mass enhancement and the analytic model, the methodological
  details, and the supplemental figures}, which includes Refs.~\cite{
  perez15,QE2017,wannier90,epw,ca80,pz81,hama13,sg15,Lejaeghere2016,Setten2018,gcl07,
  rycr09,fvg15,vg15,Sjakste2015,Marronnier2017,kara15,wrig16,Perez2017,Seeger2004,
  Grimvall1981,Galkowski2016,froe54}.}\BibitemShut {Stop}%
\bibitem [{\citenamefont {Giannozzi}\ \emph {et~al.}(2017)\citenamefont
  {Giannozzi}, \citenamefont {Andreussi}, \citenamefont {Brumme}, \citenamefont
  {Bunau}, \citenamefont {Nardelli}, \citenamefont {Calandra}, \citenamefont
  {Car}, \citenamefont {Cavazzoni}, \citenamefont {Ceresoli}, \citenamefont
  {Cococcioni}, \citenamefont {Colonna}, \citenamefont {Carnimeo},
  \citenamefont {Corso}, \citenamefont {de~Gironcoli}, \citenamefont {Delugas},
  \citenamefont {DiStasio}, \citenamefont {Ferretti}, \citenamefont {Floris},
  \citenamefont {Fratesi}, \citenamefont {Fugallo}, \citenamefont {Gebauer},
  \citenamefont {Gerstmann}, \citenamefont {Giustino}, \citenamefont {Gorni},
  \citenamefont {Jia}, \citenamefont {Kawamura}, \citenamefont {Ko},
  \citenamefont {Kokalj}, \citenamefont {K\"u\c{c}\"ukbenli}, \citenamefont
  {Lazzeri}, \citenamefont {Marsili}, \citenamefont {Marzari}, \citenamefont
  {Mauri}, \citenamefont {Nguyen}, \citenamefont {Nguyen}, \citenamefont {de-la
  Roza}, \citenamefont {Paulatto}, \citenamefont {Ponc\'e}, \citenamefont
  {Rocca}, \citenamefont {Sabatini}, \citenamefont {Santra}, \citenamefont
  {Schlipf}, \citenamefont {Seitsonen}, \citenamefont {Smogunov}, \citenamefont
  {Timrov}, \citenamefont {Thonhauser}, \citenamefont {Umari}, \citenamefont
  {Vast}, \citenamefont {Wu},\ and\ \citenamefont {Baroni}}]{QE2017}%
  \BibitemOpen
  \bibfield  {author} {\bibinfo {author} {\bibfnamefont {P.}~\bibnamefont
  {Giannozzi}}, \bibinfo {author} {\bibfnamefont {O.}~\bibnamefont
  {Andreussi}}, \bibinfo {author} {\bibfnamefont {T.}~\bibnamefont {Brumme}},
  \bibinfo {author} {\bibfnamefont {O.}~\bibnamefont {Bunau}}, \bibinfo
  {author} {\bibfnamefont {M.~B.}\ \bibnamefont {Nardelli}}, \bibinfo {author}
  {\bibfnamefont {M.}~\bibnamefont {Calandra}}, \bibinfo {author}
  {\bibfnamefont {R.}~\bibnamefont {Car}}, \bibinfo {author} {\bibfnamefont
  {C.}~\bibnamefont {Cavazzoni}}, \bibinfo {author} {\bibfnamefont
  {D.}~\bibnamefont {Ceresoli}}, \bibinfo {author} {\bibfnamefont
  {M.}~\bibnamefont {Cococcioni}}, \bibinfo {author} {\bibfnamefont
  {N.}~\bibnamefont {Colonna}}, \bibinfo {author} {\bibfnamefont
  {I.}~\bibnamefont {Carnimeo}}, \bibinfo {author} {\bibfnamefont {A.~D.}\
  \bibnamefont {Corso}}, \bibinfo {author} {\bibfnamefont {S.}~\bibnamefont
  {de~Gironcoli}}, \bibinfo {author} {\bibfnamefont {P.}~\bibnamefont
  {Delugas}}, \bibinfo {author} {\bibfnamefont {R.}~\bibnamefont {DiStasio}},
  \bibinfo {author} {\bibfnamefont {A.}~\bibnamefont {Ferretti}}, \bibinfo
  {author} {\bibfnamefont {A.}~\bibnamefont {Floris}}, \bibinfo {author}
  {\bibfnamefont {G.}~\bibnamefont {Fratesi}}, \bibinfo {author} {\bibfnamefont
  {G.}~\bibnamefont {Fugallo}}, \bibinfo {author} {\bibfnamefont
  {R.}~\bibnamefont {Gebauer}}, \bibinfo {author} {\bibfnamefont
  {U.}~\bibnamefont {Gerstmann}}, \bibinfo {author} {\bibfnamefont
  {F.}~\bibnamefont {Giustino}}, \bibinfo {author} {\bibfnamefont
  {T.}~\bibnamefont {Gorni}}, \bibinfo {author} {\bibfnamefont
  {J.}~\bibnamefont {Jia}}, \bibinfo {author} {\bibfnamefont {M.}~\bibnamefont
  {Kawamura}}, \bibinfo {author} {\bibfnamefont {H.-Y.}\ \bibnamefont {Ko}},
  \bibinfo {author} {\bibfnamefont {A.}~\bibnamefont {Kokalj}}, \bibinfo
  {author} {\bibfnamefont {E.}~\bibnamefont {K\"u\c{c}\"ukbenli}}, \bibinfo
  {author} {\bibfnamefont {M.}~\bibnamefont {Lazzeri}}, \bibinfo {author}
  {\bibfnamefont {M.}~\bibnamefont {Marsili}}, \bibinfo {author} {\bibfnamefont
  {N.}~\bibnamefont {Marzari}}, \bibinfo {author} {\bibfnamefont
  {F.}~\bibnamefont {Mauri}}, \bibinfo {author} {\bibfnamefont {N.~L.}\
  \bibnamefont {Nguyen}}, \bibinfo {author} {\bibfnamefont {H.-V.}\
  \bibnamefont {Nguyen}}, \bibinfo {author} {\bibfnamefont {A.~O.}\
  \bibnamefont {de-la Roza}}, \bibinfo {author} {\bibfnamefont
  {L.}~\bibnamefont {Paulatto}}, \bibinfo {author} {\bibfnamefont
  {S.}~\bibnamefont {Ponc\'e}}, \bibinfo {author} {\bibfnamefont
  {D.}~\bibnamefont {Rocca}}, \bibinfo {author} {\bibfnamefont
  {R.}~\bibnamefont {Sabatini}}, \bibinfo {author} {\bibfnamefont
  {B.}~\bibnamefont {Santra}}, \bibinfo {author} {\bibfnamefont
  {M.}~\bibnamefont {Schlipf}}, \bibinfo {author} {\bibfnamefont {A.~P.}\
  \bibnamefont {Seitsonen}}, \bibinfo {author} {\bibfnamefont {A.}~\bibnamefont
  {Smogunov}}, \bibinfo {author} {\bibfnamefont {I.}~\bibnamefont {Timrov}},
  \bibinfo {author} {\bibfnamefont {T.}~\bibnamefont {Thonhauser}}, \bibinfo
  {author} {\bibfnamefont {P.}~\bibnamefont {Umari}}, \bibinfo {author}
  {\bibfnamefont {N.}~\bibnamefont {Vast}}, \bibinfo {author} {\bibfnamefont
  {X.}~\bibnamefont {Wu}}, \ and\ \bibinfo {author} {\bibfnamefont
  {S.}~\bibnamefont {Baroni}},\ }\href
  {http://iopscience.iop.org/10.1088/1361-648X/aa8f79} {\bibfield  {journal}
  {\bibinfo  {journal} {J. Phys.: Condens. Matter}\ }\textbf {\bibinfo {volume}
  {29}},\ \bibinfo {pages} {465901} (\bibinfo {year} {2017})}\BibitemShut
  {NoStop}%
\bibitem [{\citenamefont {Mostofi}\ \emph {et~al.}(2014)\citenamefont
  {Mostofi}, \citenamefont {Yates}, \citenamefont {Pizzi}, \citenamefont {Lee},
  \citenamefont {Souza}, \citenamefont {Vanderbilt},\ and\ \citenamefont
  {Marzari}}]{wannier90}%
  \BibitemOpen
  \bibfield  {author} {\bibinfo {author} {\bibfnamefont {A.~A.}\ \bibnamefont
  {Mostofi}}, \bibinfo {author} {\bibfnamefont {J.~R.}\ \bibnamefont {Yates}},
  \bibinfo {author} {\bibfnamefont {G.}~\bibnamefont {Pizzi}}, \bibinfo
  {author} {\bibfnamefont {Y.-S.}\ \bibnamefont {Lee}}, \bibinfo {author}
  {\bibfnamefont {I.}~\bibnamefont {Souza}}, \bibinfo {author} {\bibfnamefont
  {D.}~\bibnamefont {Vanderbilt}}, \ and\ \bibinfo {author} {\bibfnamefont
  {N.}~\bibnamefont {Marzari}},\ }\href {\doibase
  http://dx.doi.org/10.1016/j.cpc.2014.05.003} {\bibfield  {journal} {\bibinfo
  {journal} {Comput. Phys. Commun.}\ }\textbf {\bibinfo {volume} {185}},\
  \bibinfo {pages} {2309 } (\bibinfo {year} {2014})}\BibitemShut {NoStop}%
\bibitem [{\citenamefont {Ponc\'e}\ \emph {et~al.}(2016)\citenamefont
  {Ponc\'e}, \citenamefont {Margine}, \citenamefont {Verdi},\ and\
  \citenamefont {Giustino}}]{epw}%
  \BibitemOpen
  \bibfield  {author} {\bibinfo {author} {\bibfnamefont {S.}~\bibnamefont
  {Ponc\'e}}, \bibinfo {author} {\bibfnamefont {E.}~\bibnamefont {Margine}},
  \bibinfo {author} {\bibfnamefont {C.}~\bibnamefont {Verdi}}, \ and\ \bibinfo
  {author} {\bibfnamefont {F.}~\bibnamefont {Giustino}},\ }\href {\doibase
  http://dx.doi.org/10.1016/j.cpc.2016.07.028} {\bibfield  {journal} {\bibinfo
  {journal} {Comput. Phys. Commun.}\ }\textbf {\bibinfo {volume} {209}},\
  \bibinfo {pages} {116 } (\bibinfo {year} {2016})}\BibitemShut {NoStop}%
\bibitem [{\citenamefont {Ceperley}\ and\ \citenamefont {Alder}(1980)}]{ca80}%
  \BibitemOpen
  \bibfield  {author} {\bibinfo {author} {\bibfnamefont {D.~M.}\ \bibnamefont
  {Ceperley}}\ and\ \bibinfo {author} {\bibfnamefont {B.~J.}\ \bibnamefont
  {Alder}},\ }\href {\doibase 10.1103/PhysRevLett.45.566} {\bibfield  {journal}
  {\bibinfo  {journal} {Phys. Rev. Lett.}\ }\textbf {\bibinfo {volume} {45}},\
  \bibinfo {pages} {566} (\bibinfo {year} {1980})}\BibitemShut {NoStop}%
\bibitem [{\citenamefont {Perdew}\ and\ \citenamefont {Zunger}(1981)}]{pz81}%
  \BibitemOpen
  \bibfield  {author} {\bibinfo {author} {\bibfnamefont {J.~P.}\ \bibnamefont
  {Perdew}}\ and\ \bibinfo {author} {\bibfnamefont {A.}~\bibnamefont
  {Zunger}},\ }\href {\doibase 10.1103/PhysRevB.23.5048} {\bibfield  {journal}
  {\bibinfo  {journal} {Phys. Rev. B}\ }\textbf {\bibinfo {volume} {23}},\
  \bibinfo {pages} {5048} (\bibinfo {year} {1981})}\BibitemShut {NoStop}%
\bibitem [{\citenamefont {Hamann}(2013)}]{hama13}%
  \BibitemOpen
  \bibfield  {author} {\bibinfo {author} {\bibfnamefont {D.~R.}\ \bibnamefont
  {Hamann}},\ }\href {\doibase 10.1103/PhysRevB.88.085117} {\bibfield
  {journal} {\bibinfo  {journal} {Phys. Rev. B}\ }\textbf {\bibinfo {volume}
  {88}},\ \bibinfo {pages} {085117} (\bibinfo {year} {2013})}\BibitemShut
  {NoStop}%
\bibitem [{\citenamefont {Schlipf}\ and\ \citenamefont {Gygi}(2015)}]{sg15}%
  \BibitemOpen
  \bibfield  {author} {\bibinfo {author} {\bibfnamefont {M.}~\bibnamefont
  {Schlipf}}\ and\ \bibinfo {author} {\bibfnamefont {F.}~\bibnamefont {Gygi}},\
  }\href {\doibase http://dx.doi.org/10.1016/j.cpc.2015.05.011} {\bibfield
  {journal} {\bibinfo  {journal} {Comput. Phys. Commun.}\ }\textbf {\bibinfo
  {volume} {196}},\ \bibinfo {pages} {36} (\bibinfo {year} {2015})}\BibitemShut
  {NoStop}%
\bibitem [{\citenamefont {Lejaeghere}\ \emph {et~al.}(2016)\citenamefont
  {Lejaeghere}, \citenamefont {Bihlmayer}, \citenamefont {Bj{\"o}rkman},
  \citenamefont {Blaha}, \citenamefont {Bl{\"u}gel}, \citenamefont {Blum},
  \citenamefont {Caliste}, \citenamefont {Castelli}, \citenamefont {Clark},
  \citenamefont {Dal~Corso}, \citenamefont {de~Gironcoli}, \citenamefont
  {Deutsch}, \citenamefont {Dewhurst}, \citenamefont {Di~Marco}, \citenamefont
  {Draxl}, \citenamefont {Du{\l}ak}, \citenamefont {Eriksson}, \citenamefont
  {Flores-Livas}, \citenamefont {Garrity}, \citenamefont {Genovese},
  \citenamefont {Giannozzi}, \citenamefont {Giantomassi}, \citenamefont
  {Goedecker}, \citenamefont {Gonze}, \citenamefont {Gr{\r a}n{\"a}s},
  \citenamefont {Gross}, \citenamefont {Gulans}, \citenamefont {Gygi},
  \citenamefont {Hamann}, \citenamefont {Hasnip}, \citenamefont {Holzwarth},
  \citenamefont {Iu{\c s}an}, \citenamefont {Jochym}, \citenamefont {Jollet},
  \citenamefont {Jones}, \citenamefont {Kresse}, \citenamefont {Koepernik},
  \citenamefont {K{\"u}{\c c}{\"u}kbenli}, \citenamefont {Kvashnin},
  \citenamefont {Locht}, \citenamefont {Lubeck}, \citenamefont {Marsman},
  \citenamefont {Marzari}, \citenamefont {Nitzsche}, \citenamefont
  {Nordstr{\"o}m}, \citenamefont {Ozaki}, \citenamefont {Paulatto},
  \citenamefont {Pickard}, \citenamefont {Poelmans}, \citenamefont {Probert},
  \citenamefont {Refson}, \citenamefont {Richter}, \citenamefont {Rignanese},
  \citenamefont {Saha}, \citenamefont {Scheffler}, \citenamefont {Schlipf},
  \citenamefont {Schwarz}, \citenamefont {Sharma}, \citenamefont {Tavazza},
  \citenamefont {Thunstr{\"o}m}, \citenamefont {Tkatchenko}, \citenamefont
  {Torrent}, \citenamefont {Vanderbilt}, \citenamefont {van Setten},
  \citenamefont {Van~Speybroeck}, \citenamefont {Wills}, \citenamefont {Yates},
  \citenamefont {Zhang},\ and\ \citenamefont {Cottenier}}]{Lejaeghere2016}%
  \BibitemOpen
  \bibfield  {author} {\bibinfo {author} {\bibfnamefont {K.}~\bibnamefont
  {Lejaeghere}}, \bibinfo {author} {\bibfnamefont {G.}~\bibnamefont
  {Bihlmayer}}, \bibinfo {author} {\bibfnamefont {T.}~\bibnamefont
  {Bj{\"o}rkman}}, \bibinfo {author} {\bibfnamefont {P.}~\bibnamefont {Blaha}},
  \bibinfo {author} {\bibfnamefont {S.}~\bibnamefont {Bl{\"u}gel}}, \bibinfo
  {author} {\bibfnamefont {V.}~\bibnamefont {Blum}}, \bibinfo {author}
  {\bibfnamefont {D.}~\bibnamefont {Caliste}}, \bibinfo {author} {\bibfnamefont
  {I.~E.}\ \bibnamefont {Castelli}}, \bibinfo {author} {\bibfnamefont {S.~J.}\
  \bibnamefont {Clark}}, \bibinfo {author} {\bibfnamefont {A.}~\bibnamefont
  {Dal~Corso}}, \bibinfo {author} {\bibfnamefont {S.}~\bibnamefont
  {de~Gironcoli}}, \bibinfo {author} {\bibfnamefont {T.}~\bibnamefont
  {Deutsch}}, \bibinfo {author} {\bibfnamefont {J.~K.}\ \bibnamefont
  {Dewhurst}}, \bibinfo {author} {\bibfnamefont {I.}~\bibnamefont {Di~Marco}},
  \bibinfo {author} {\bibfnamefont {C.}~\bibnamefont {Draxl}}, \bibinfo
  {author} {\bibfnamefont {M.}~\bibnamefont {Du{\l}ak}}, \bibinfo {author}
  {\bibfnamefont {O.}~\bibnamefont {Eriksson}}, \bibinfo {author}
  {\bibfnamefont {J.~A.}\ \bibnamefont {Flores-Livas}}, \bibinfo {author}
  {\bibfnamefont {K.~F.}\ \bibnamefont {Garrity}}, \bibinfo {author}
  {\bibfnamefont {L.}~\bibnamefont {Genovese}}, \bibinfo {author}
  {\bibfnamefont {P.}~\bibnamefont {Giannozzi}}, \bibinfo {author}
  {\bibfnamefont {M.}~\bibnamefont {Giantomassi}}, \bibinfo {author}
  {\bibfnamefont {S.}~\bibnamefont {Goedecker}}, \bibinfo {author}
  {\bibfnamefont {X.}~\bibnamefont {Gonze}}, \bibinfo {author} {\bibfnamefont
  {O.}~\bibnamefont {Gr{\r a}n{\"a}s}}, \bibinfo {author} {\bibfnamefont
  {E.~K.~U.}\ \bibnamefont {Gross}}, \bibinfo {author} {\bibfnamefont
  {A.}~\bibnamefont {Gulans}}, \bibinfo {author} {\bibfnamefont
  {F.}~\bibnamefont {Gygi}}, \bibinfo {author} {\bibfnamefont {D.~R.}\
  \bibnamefont {Hamann}}, \bibinfo {author} {\bibfnamefont {P.~J.}\
  \bibnamefont {Hasnip}}, \bibinfo {author} {\bibfnamefont {N.~A.~W.}\
  \bibnamefont {Holzwarth}}, \bibinfo {author} {\bibfnamefont {D.}~\bibnamefont
  {Iu{\c s}an}}, \bibinfo {author} {\bibfnamefont {D.~B.}\ \bibnamefont
  {Jochym}}, \bibinfo {author} {\bibfnamefont {F.}~\bibnamefont {Jollet}},
  \bibinfo {author} {\bibfnamefont {D.}~\bibnamefont {Jones}}, \bibinfo
  {author} {\bibfnamefont {G.}~\bibnamefont {Kresse}}, \bibinfo {author}
  {\bibfnamefont {K.}~\bibnamefont {Koepernik}}, \bibinfo {author}
  {\bibfnamefont {E.}~\bibnamefont {K{\"u}{\c c}{\"u}kbenli}}, \bibinfo
  {author} {\bibfnamefont {Y.~O.}\ \bibnamefont {Kvashnin}}, \bibinfo {author}
  {\bibfnamefont {I.~L.~M.}\ \bibnamefont {Locht}}, \bibinfo {author}
  {\bibfnamefont {S.}~\bibnamefont {Lubeck}}, \bibinfo {author} {\bibfnamefont
  {M.}~\bibnamefont {Marsman}}, \bibinfo {author} {\bibfnamefont
  {N.}~\bibnamefont {Marzari}}, \bibinfo {author} {\bibfnamefont
  {U.}~\bibnamefont {Nitzsche}}, \bibinfo {author} {\bibfnamefont
  {L.}~\bibnamefont {Nordstr{\"o}m}}, \bibinfo {author} {\bibfnamefont
  {T.}~\bibnamefont {Ozaki}}, \bibinfo {author} {\bibfnamefont
  {L.}~\bibnamefont {Paulatto}}, \bibinfo {author} {\bibfnamefont {C.~J.}\
  \bibnamefont {Pickard}}, \bibinfo {author} {\bibfnamefont {W.}~\bibnamefont
  {Poelmans}}, \bibinfo {author} {\bibfnamefont {M.~I.~J.}\ \bibnamefont
  {Probert}}, \bibinfo {author} {\bibfnamefont {K.}~\bibnamefont {Refson}},
  \bibinfo {author} {\bibfnamefont {M.}~\bibnamefont {Richter}}, \bibinfo
  {author} {\bibfnamefont {G.-M.}\ \bibnamefont {Rignanese}}, \bibinfo {author}
  {\bibfnamefont {S.}~\bibnamefont {Saha}}, \bibinfo {author} {\bibfnamefont
  {M.}~\bibnamefont {Scheffler}}, \bibinfo {author} {\bibfnamefont
  {M.}~\bibnamefont {Schlipf}}, \bibinfo {author} {\bibfnamefont
  {K.}~\bibnamefont {Schwarz}}, \bibinfo {author} {\bibfnamefont
  {S.}~\bibnamefont {Sharma}}, \bibinfo {author} {\bibfnamefont
  {F.}~\bibnamefont {Tavazza}}, \bibinfo {author} {\bibfnamefont
  {P.}~\bibnamefont {Thunstr{\"o}m}}, \bibinfo {author} {\bibfnamefont
  {A.}~\bibnamefont {Tkatchenko}}, \bibinfo {author} {\bibfnamefont
  {M.}~\bibnamefont {Torrent}}, \bibinfo {author} {\bibfnamefont
  {D.}~\bibnamefont {Vanderbilt}}, \bibinfo {author} {\bibfnamefont {M.~J.}\
  \bibnamefont {van Setten}}, \bibinfo {author} {\bibfnamefont
  {V.}~\bibnamefont {Van~Speybroeck}}, \bibinfo {author} {\bibfnamefont
  {J.~M.}\ \bibnamefont {Wills}}, \bibinfo {author} {\bibfnamefont {J.~R.}\
  \bibnamefont {Yates}}, \bibinfo {author} {\bibfnamefont {G.-X.}\ \bibnamefont
  {Zhang}}, \ and\ \bibinfo {author} {\bibfnamefont {S.}~\bibnamefont
  {Cottenier}},\ }\href {\doibase 10.1126/science.aad3000} {\bibfield
  {journal} {\bibinfo  {journal} {Science}\ }\textbf {\bibinfo {volume}
  {351}},\ \bibinfo {pages} {aad3000} (\bibinfo {year} {2016})}\BibitemShut
  {NoStop}%
\bibitem [{\citenamefont {van Setten}\ \emph {et~al.}(2018)\citenamefont {van
  Setten}, \citenamefont {Giantomassi}, \citenamefont {Bousquet}, \citenamefont
  {Verstraete}, \citenamefont {Hamann}, \citenamefont {Gonze},\ and\
  \citenamefont {Rignanese}}]{Setten2018}%
  \BibitemOpen
  \bibfield  {author} {\bibinfo {author} {\bibfnamefont {M.}~\bibnamefont {van
  Setten}}, \bibinfo {author} {\bibfnamefont {M.}~\bibnamefont {Giantomassi}},
  \bibinfo {author} {\bibfnamefont {E.}~\bibnamefont {Bousquet}}, \bibinfo
  {author} {\bibfnamefont {M.}~\bibnamefont {Verstraete}}, \bibinfo {author}
  {\bibfnamefont {D.}~\bibnamefont {Hamann}}, \bibinfo {author} {\bibfnamefont
  {X.}~\bibnamefont {Gonze}}, \ and\ \bibinfo {author} {\bibfnamefont {G.-M.}\
  \bibnamefont {Rignanese}},\ }\href {\doibase
  https://doi.org/10.1016/j.cpc.2018.01.012} {\bibfield  {journal} {\bibinfo
  {journal} {Comput. Phys. Commun.}\ }\textbf {\bibinfo {volume} {226}},\
  \bibinfo {pages} {39 } (\bibinfo {year} {2018})}\BibitemShut {NoStop}%
\bibitem [{\citenamefont {Giustino}\ \emph {et~al.}(2007)\citenamefont
  {Giustino}, \citenamefont {Cohen},\ and\ \citenamefont {Louie}}]{gcl07}%
  \BibitemOpen
  \bibfield  {author} {\bibinfo {author} {\bibfnamefont {F.}~\bibnamefont
  {Giustino}}, \bibinfo {author} {\bibfnamefont {M.~L.}\ \bibnamefont {Cohen}},
  \ and\ \bibinfo {author} {\bibfnamefont {S.~G.}\ \bibnamefont {Louie}},\
  }\href {\doibase 10.1103/PhysRevB.76.165108} {\bibfield  {journal} {\bibinfo
  {journal} {Phys. Rev. B}\ }\textbf {\bibinfo {volume} {76}},\ \bibinfo
  {pages} {165108} (\bibinfo {year} {2007})}\BibitemShut {NoStop}%
\bibitem [{\citenamefont {Rycroft}(2009)}]{rycr09}%
  \BibitemOpen
  \bibfield  {author} {\bibinfo {author} {\bibfnamefont {C.~H.}\ \bibnamefont
  {Rycroft}},\ }\href {\doibase 10.1063/1.3215722} {\bibfield  {journal}
  {\bibinfo  {journal} {Chaos}\ }\textbf {\bibinfo {volume} {19}},\ \bibinfo
  {pages} {041111} (\bibinfo {year} {2009})}\BibitemShut {NoStop}%
\bibitem [{\citenamefont {Filip}\ \emph {et~al.}(2015)\citenamefont {Filip},
  \citenamefont {Verdi},\ and\ \citenamefont {Giustino}}]{fvg15}%
  \BibitemOpen
  \bibfield  {author} {\bibinfo {author} {\bibfnamefont {M.~R.}\ \bibnamefont
  {Filip}}, \bibinfo {author} {\bibfnamefont {C.}~\bibnamefont {Verdi}}, \ and\
  \bibinfo {author} {\bibfnamefont {F.}~\bibnamefont {Giustino}},\ }\href
  {\doibase 10.1021/acs.jpcc.5b07891} {\bibfield  {journal} {\bibinfo
  {journal} {J. Phys. Chem. C}\ }\textbf {\bibinfo {volume} {119}},\ \bibinfo
  {pages} {25209} (\bibinfo {year} {2015})}\BibitemShut {NoStop}%
\bibitem [{\citenamefont {Verdi}\ and\ \citenamefont {Giustino}(2015)}]{vg15}%
  \BibitemOpen
  \bibfield  {author} {\bibinfo {author} {\bibfnamefont {C.}~\bibnamefont
  {Verdi}}\ and\ \bibinfo {author} {\bibfnamefont {F.}~\bibnamefont
  {Giustino}},\ }\href {\doibase 10.1103/PhysRevLett.115.176401} {\bibfield
  {journal} {\bibinfo  {journal} {Phys. Rev. Lett.}\ }\textbf {\bibinfo
  {volume} {115}},\ \bibinfo {pages} {176401} (\bibinfo {year}
  {2015})}\BibitemShut {NoStop}%
\bibitem [{\citenamefont {Sjakste}\ \emph {et~al.}(2015)\citenamefont
  {Sjakste}, \citenamefont {Vast}, \citenamefont {Calandra},\ and\
  \citenamefont {Mauri}}]{Sjakste2015}%
  \BibitemOpen
  \bibfield  {author} {\bibinfo {author} {\bibfnamefont {J.}~\bibnamefont
  {Sjakste}}, \bibinfo {author} {\bibfnamefont {N.}~\bibnamefont {Vast}},
  \bibinfo {author} {\bibfnamefont {M.}~\bibnamefont {Calandra}}, \ and\
  \bibinfo {author} {\bibfnamefont {F.}~\bibnamefont {Mauri}},\ }\href
  {\doibase 10.1103/PhysRevB.92.054307} {\bibfield  {journal} {\bibinfo
  {journal} {Phys. Rev. B}\ }\textbf {\bibinfo {volume} {92}},\ \bibinfo
  {pages} {054307} (\bibinfo {year} {2015})}\BibitemShut {NoStop}%
\bibitem [{\citenamefont {Marronnier}\ \emph {et~al.}(2017)\citenamefont
  {Marronnier}, \citenamefont {Lee}, \citenamefont {Geffroy}, \citenamefont
  {Even}, \citenamefont {Bonnassieux},\ and\ \citenamefont
  {Roma}}]{Marronnier2017}%
  \BibitemOpen
  \bibfield  {author} {\bibinfo {author} {\bibfnamefont {A.}~\bibnamefont
  {Marronnier}}, \bibinfo {author} {\bibfnamefont {H.}~\bibnamefont {Lee}},
  \bibinfo {author} {\bibfnamefont {B.}~\bibnamefont {Geffroy}}, \bibinfo
  {author} {\bibfnamefont {J.}~\bibnamefont {Even}}, \bibinfo {author}
  {\bibfnamefont {Y.}~\bibnamefont {Bonnassieux}}, \ and\ \bibinfo {author}
  {\bibfnamefont {G.}~\bibnamefont {Roma}},\ }\href {\doibase
  10.1021/acs.jpclett.7b00807} {\bibfield  {journal} {\bibinfo  {journal} {J.
  Phys. Chem. Lett.}\ }\textbf {\bibinfo {volume} {8}},\ \bibinfo {pages}
  {2659} (\bibinfo {year} {2017})}\BibitemShut {NoStop}%
\bibitem [{\citenamefont {P\'erez-Osorio}\ \emph {et~al.}(2017)\citenamefont
  {P\'erez-Osorio}, \citenamefont {Champagne}, \citenamefont {Zacharias},
  \citenamefont {Rignanese},\ and\ \citenamefont {Giustino}}]{Perez2017}%
  \BibitemOpen
  \bibfield  {author} {\bibinfo {author} {\bibfnamefont {M.~A.}\ \bibnamefont
  {P\'erez-Osorio}}, \bibinfo {author} {\bibfnamefont {A.}~\bibnamefont
  {Champagne}}, \bibinfo {author} {\bibfnamefont {M.}~\bibnamefont
  {Zacharias}}, \bibinfo {author} {\bibfnamefont {G.-M.}\ \bibnamefont
  {Rignanese}}, \ and\ \bibinfo {author} {\bibfnamefont {F.}~\bibnamefont
  {Giustino}},\ }\href {\doibase 10.1021/acs.jpcc.7b07121} {\bibfield
  {journal} {\bibinfo  {journal} {J. Phys. Chem. C}\ }\textbf {\bibinfo
  {volume} {121}},\ \bibinfo {pages} {18459} (\bibinfo {year}
  {2017})}\BibitemShut {NoStop}%
\bibitem [{\citenamefont {Seeger}(2004)}]{Seeger2004}%
  \BibitemOpen
  \bibfield  {author} {\bibinfo {author} {\bibfnamefont {K.}~\bibnamefont
  {Seeger}},\ }\href {https://books.google.co.uk/books?id=il4nyDF0IJIC} {\emph
  {\bibinfo {title} {Semiconductor Physics}}}\ (\bibinfo  {publisher}
  {Springer},\ \bibinfo {address} {New York},\ \bibinfo {year}
  {2004})\BibitemShut {NoStop}%
\bibitem [{\citenamefont {Grimvall}(1981)}]{Grimvall1981}%
  \BibitemOpen
  \bibfield  {author} {\bibinfo {author} {\bibfnamefont {G.}~\bibnamefont
  {Grimvall}},\ }\href {http://openlibrary.org/books/OL4255043M} {\emph
  {\bibinfo {title} {The Electron-Phonon Interaction in Metals}}}\ (\bibinfo
  {publisher} {North Holland Publishing Co.},\ \bibinfo {year}
  {1981})\BibitemShut {NoStop}%
\bibitem [{\citenamefont {Galkowski}\ \emph {et~al.}(2016)\citenamefont
  {Galkowski}, \citenamefont {Mitioglu}, \citenamefont {Miyata}, \citenamefont
  {Plochocka}, \citenamefont {Portugall}, \citenamefont {Eperon}, \citenamefont
  {Wang}, \citenamefont {Stergiopoulos}, \citenamefont {Stranks}, \citenamefont
  {Snaith},\ and\ \citenamefont {Nicholas}}]{Galkowski2016}%
  \BibitemOpen
  \bibfield  {author} {\bibinfo {author} {\bibfnamefont {K.}~\bibnamefont
  {Galkowski}}, \bibinfo {author} {\bibfnamefont {A.}~\bibnamefont {Mitioglu}},
  \bibinfo {author} {\bibfnamefont {A.}~\bibnamefont {Miyata}}, \bibinfo
  {author} {\bibfnamefont {P.}~\bibnamefont {Plochocka}}, \bibinfo {author}
  {\bibfnamefont {O.}~\bibnamefont {Portugall}}, \bibinfo {author}
  {\bibfnamefont {G.~E.}\ \bibnamefont {Eperon}}, \bibinfo {author}
  {\bibfnamefont {J.~T.-W.}\ \bibnamefont {Wang}}, \bibinfo {author}
  {\bibfnamefont {T.}~\bibnamefont {Stergiopoulos}}, \bibinfo {author}
  {\bibfnamefont {S.~D.}\ \bibnamefont {Stranks}}, \bibinfo {author}
  {\bibfnamefont {H.~J.}\ \bibnamefont {Snaith}}, \ and\ \bibinfo {author}
  {\bibfnamefont {R.~J.}\ \bibnamefont {Nicholas}},\ }\href {\doibase
  10.1039/C5EE03435C} {\bibfield  {journal} {\bibinfo  {journal} {Energy
  Environ. Sci.}\ }\textbf {\bibinfo {volume} {9}},\ \bibinfo {pages} {962}
  (\bibinfo {year} {2016})}\BibitemShut {NoStop}%
\bibitem [{\citenamefont {Fr\"ohlich}(1954)}]{froe54}%
  \BibitemOpen
  \bibfield  {author} {\bibinfo {author} {\bibfnamefont {H.}~\bibnamefont
  {Fr\"ohlich}},\ }\href {\doibase 10.1080/00018735400101213} {\bibfield
  {journal} {\bibinfo  {journal} {Adv. Phys.}\ }\textbf {\bibinfo {volume}
  {3}},\ \bibinfo {pages} {325} (\bibinfo {year} {1954})}\BibitemShut {NoStop}%
\bibitem [{\citenamefont {Even}\ \emph {et~al.}(2013)\citenamefont {Even},
  \citenamefont {Pedesseau}, \citenamefont {Jancu},\ and\ \citenamefont
  {Katan}}]{epjk13}%
  \BibitemOpen
  \bibfield  {author} {\bibinfo {author} {\bibfnamefont {J.}~\bibnamefont
  {Even}}, \bibinfo {author} {\bibfnamefont {L.}~\bibnamefont {Pedesseau}},
  \bibinfo {author} {\bibfnamefont {J.-M.}\ \bibnamefont {Jancu}}, \ and\
  \bibinfo {author} {\bibfnamefont {C.}~\bibnamefont {Katan}},\ }\href
  {\doibase 10.1021/jz401532q} {\bibfield  {journal} {\bibinfo  {journal} {J.
  Phys. Chem. Lett.}\ }\textbf {\bibinfo {volume} {4}},\ \bibinfo {pages}
  {2999} (\bibinfo {year} {2013})}\BibitemShut {NoStop}%
\bibitem [{\citenamefont {Umari}\ \emph {et~al.}(2014)\citenamefont {Umari},
  \citenamefont {Mosconi},\ and\ \citenamefont {De~Angelis}}]{umd14}%
  \BibitemOpen
  \bibfield  {author} {\bibinfo {author} {\bibfnamefont {P.}~\bibnamefont
  {Umari}}, \bibinfo {author} {\bibfnamefont {E.}~\bibnamefont {Mosconi}}, \
  and\ \bibinfo {author} {\bibfnamefont {F.}~\bibnamefont {De~Angelis}},\
  }\href {http://dx.doi.org/10.1038/srep04467} {\bibfield  {journal} {\bibinfo
  {journal} {Sci. Rep.}\ }\textbf {\bibinfo {volume} {4}},\ \bibinfo {pages}
  {4467} (\bibinfo {year} {2014})}\BibitemShut {NoStop}%
\bibitem [{\citenamefont {Filip}\ and\ \citenamefont {Giustino}(2014)}]{fg14}%
  \BibitemOpen
  \bibfield  {author} {\bibinfo {author} {\bibfnamefont {M.~R.}\ \bibnamefont
  {Filip}}\ and\ \bibinfo {author} {\bibfnamefont {F.}~\bibnamefont
  {Giustino}},\ }\href {\doibase 10.1103/PhysRevB.90.245145} {\bibfield
  {journal} {\bibinfo  {journal} {Phys. Rev. B}\ }\textbf {\bibinfo {volume}
  {90}},\ \bibinfo {pages} {245145} (\bibinfo {year} {2014})}\BibitemShut
  {NoStop}%
\bibitem [{\citenamefont {Kawai}\ \emph {et~al.}(2015)\citenamefont {Kawai},
  \citenamefont {Giorgi}, \citenamefont {Marini},\ and\ \citenamefont
  {Yamashita}}]{Kawai2015}%
  \BibitemOpen
  \bibfield  {author} {\bibinfo {author} {\bibfnamefont {H.}~\bibnamefont
  {Kawai}}, \bibinfo {author} {\bibfnamefont {G.}~\bibnamefont {Giorgi}},
  \bibinfo {author} {\bibfnamefont {A.}~\bibnamefont {Marini}}, \ and\ \bibinfo
  {author} {\bibfnamefont {K.}~\bibnamefont {Yamashita}},\ }\href {\doibase
  10.1021/acs.nanolett.5b00109} {\bibfield  {journal} {\bibinfo  {journal}
  {Nano Lett.}\ }\textbf {\bibinfo {volume} {15}},\ \bibinfo {pages} {3103}
  (\bibinfo {year} {2015})}\BibitemShut {NoStop}%
\bibitem [{\citenamefont {Saidi}\ \emph {et~al.}(2016)\citenamefont {Saidi},
  \citenamefont {Ponc\'e},\ and\ \citenamefont {Monserrat}}]{spm16}%
  \BibitemOpen
  \bibfield  {author} {\bibinfo {author} {\bibfnamefont {W.~A.}\ \bibnamefont
  {Saidi}}, \bibinfo {author} {\bibfnamefont {S.}~\bibnamefont {Ponc\'e}}, \
  and\ \bibinfo {author} {\bibfnamefont {B.}~\bibnamefont {Monserrat}},\ }\href
  {\doibase 10.1021/acs.jpclett.6b02560} {\bibfield  {journal} {\bibinfo
  {journal} {J. Phys. Chem. Lett.}\ }\textbf {\bibinfo {volume} {7}},\ \bibinfo
  {pages} {5247} (\bibinfo {year} {2016})}\BibitemShut {NoStop}%
\bibitem [{\citenamefont {Bokdam}\ \emph {et~al.}(2016)\citenamefont {Bokdam},
  \citenamefont {Sander}, \citenamefont {Stroppa}, \citenamefont {Picozzi},
  \citenamefont {Sarma}, \citenamefont {Franchini},\ and\ \citenamefont
  {Kresse}}]{Bokdam2016}%
  \BibitemOpen
  \bibfield  {author} {\bibinfo {author} {\bibfnamefont {M.}~\bibnamefont
  {Bokdam}}, \bibinfo {author} {\bibfnamefont {T.}~\bibnamefont {Sander}},
  \bibinfo {author} {\bibfnamefont {A.}~\bibnamefont {Stroppa}}, \bibinfo
  {author} {\bibfnamefont {S.}~\bibnamefont {Picozzi}}, \bibinfo {author}
  {\bibfnamefont {D.~D.}\ \bibnamefont {Sarma}}, \bibinfo {author}
  {\bibfnamefont {C.}~\bibnamefont {Franchini}}, \ and\ \bibinfo {author}
  {\bibfnamefont {G.}~\bibnamefont {Kresse}},\ }\href
  {http://dx.doi.org/10.1038/srep28618} {\bibfield  {journal} {\bibinfo
  {journal} {Sci. Rep.}\ }\textbf {\bibinfo {volume} {6}},\ \bibinfo {pages}
  {28618} (\bibinfo {year} {2016})}\BibitemShut {NoStop}%
\bibitem [{wea()}]{weak-regime}%
  \BibitemOpen
  \href@noop {} {}\bibinfo {note} {In the Fr\"ohlich model, the strength of the
  electron-phonon coupling in CH$_3$NH$_3$PbI$_3$ is $\alpha = 1.4$, so that an
  average of $\alpha/2 = 0.7$ phonons are involved in a scattering process
  \cite{Mahan2000}. {For this weak coupling the exact solution is close to the
  result obtained in Rayleigh-Schr\"odinger perturbation theory}
  \cite{Devreese2009}.}\BibitemShut {Stop}%
\bibitem [{\citenamefont {Mahan}(2000)}]{Mahan2000}%
  \BibitemOpen
  \bibfield  {author} {\bibinfo {author} {\bibfnamefont {G.}~\bibnamefont
  {Mahan}},\ }\href {https://books.google.co.uk/books?id=xzSgZ4-yyMEC} {\emph
  {\bibinfo {title} {Many-Particle Physics}}},\ Physics of Solids and Liquids\
  (\bibinfo  {publisher} {Springer US},\ \bibinfo {address} {New York},\
  \bibinfo {year} {2000})\BibitemShut {NoStop}%
\bibitem [{\citenamefont {Devreese}\ and\ \citenamefont
  {Alexandrov}(2009)}]{Devreese2009}%
  \BibitemOpen
  \bibfield  {author} {\bibinfo {author} {\bibfnamefont {J.~T.}\ \bibnamefont
  {Devreese}}\ and\ \bibinfo {author} {\bibfnamefont {A.~S.}\ \bibnamefont
  {Alexandrov}},\ }\href {http://stacks.iop.org/0034-4885/72/i=6/a=066501}
  {\bibfield  {journal} {\bibinfo  {journal} {Rep. Prog. Phys.}\ }\textbf
  {\bibinfo {volume} {72}},\ \bibinfo {pages} {066501} (\bibinfo {year}
  {2009})}\BibitemShut {NoStop}%
\bibitem [{\citenamefont {Fan}(1951)}]{fan51}%
  \BibitemOpen
  \bibfield  {author} {\bibinfo {author} {\bibfnamefont {H.~Y.}\ \bibnamefont
  {Fan}},\ }\href {\doibase 10.1103/PhysRev.82.900} {\bibfield  {journal}
  {\bibinfo  {journal} {Phys. Rev.}\ }\textbf {\bibinfo {volume} {82}},\
  \bibinfo {pages} {900} (\bibinfo {year} {1951})}\BibitemShut {NoStop}%
\bibitem [{\citenamefont {Migdal}(1958)}]{Migdal1958}%
  \BibitemOpen
  \bibfield  {author} {\bibinfo {author} {\bibfnamefont {A.~B.}\ \bibnamefont
  {Migdal}},\ }\href {http://www.jetp.ac.ru/cgi-bin/e/index/e/7/6/p996?a=list}
  {\bibfield  {journal} {\bibinfo  {journal} {Sov. Phys. JETP}\ }\textbf
  {\bibinfo {volume} {7}},\ \bibinfo {pages} {996} (\bibinfo {year}
  {1958})}\BibitemShut {NoStop}%
\bibitem [{\citenamefont {Giustino}(2017)}]{gius17}%
  \BibitemOpen
  \bibfield  {author} {\bibinfo {author} {\bibfnamefont {F.}~\bibnamefont
  {Giustino}},\ }\href {\doibase 10.1103/RevModPhys.89.015003} {\bibfield
  {journal} {\bibinfo  {journal} {Rev. Mod. Phys.}\ }\textbf {\bibinfo {volume}
  {89}},\ \bibinfo {pages} {015003} (\bibinfo {year} {2017})}\BibitemShut
  {NoStop}%
\bibitem [{\citenamefont {Hedin}\ and\ \citenamefont
  {Lundqvist}(1969)}]{Hedin1969}%
  \BibitemOpen
  \bibfield  {author} {\bibinfo {author} {\bibfnamefont {L.}~\bibnamefont
  {Hedin}}\ and\ \bibinfo {author} {\bibfnamefont {S.~O.}\ \bibnamefont
  {Lundqvist}},\ }\href {http://dx.doi.org/10.1016/S0081-1947(08)60615-3}
  {\emph {\bibinfo {title} {Effects of Electron-Electron and Electron-Phonon
  Interactions on the One-Electron States of Solids}}},\ edited by\ \bibinfo
  {editor} {\bibfnamefont {F.}~\bibnamefont {Seitz}}, \bibinfo {editor}
  {\bibfnamefont {D.}~\bibnamefont {Turnbull}}, \ and\ \bibinfo {editor}
  {\bibfnamefont {H.}~\bibnamefont {Ehrenreich}},\ \bibinfo {series} {Solid
  State Physics}, Vol.~\bibinfo {volume} {23}\ (\bibinfo  {publisher} {Academic
  Press},\ \bibinfo {address} {New York},\ \bibinfo {year} {1969})\ pp.\
  \bibinfo {pages} {1--181}\BibitemShut {NoStop}%
\bibitem [{\citenamefont {Grimvall}(1968)}]{Grimvall1968}%
  \BibitemOpen
  \bibfield  {author} {\bibinfo {author} {\bibfnamefont {G.}~\bibnamefont
  {Grimvall}},\ }\href {\doibase https://doi.org/10.1016/0022-3697(68)90214-X}
  {\bibfield  {journal} {\bibinfo  {journal} {J. Phys. Chem. Solids}\ }\textbf
  {\bibinfo {volume} {29}},\ \bibinfo {pages} {1221 } (\bibinfo {year}
  {1968})}\BibitemShut {NoStop}%
\bibitem [{\citenamefont {Miyata}\ \emph {et~al.}(2015)\citenamefont {Miyata},
  \citenamefont {Mitioglu}, \citenamefont {Plochocka}, \citenamefont
  {Portugall}, \citenamefont {Wang}, \citenamefont {Stranks}, \citenamefont
  {Snaith},\ and\ \citenamefont {Nicholas}}]{Miyata2015}%
  \BibitemOpen
  \bibfield  {author} {\bibinfo {author} {\bibfnamefont {A.}~\bibnamefont
  {Miyata}}, \bibinfo {author} {\bibfnamefont {A.}~\bibnamefont {Mitioglu}},
  \bibinfo {author} {\bibfnamefont {P.}~\bibnamefont {Plochocka}}, \bibinfo
  {author} {\bibfnamefont {O.}~\bibnamefont {Portugall}}, \bibinfo {author}
  {\bibfnamefont {J.~T.-W.}\ \bibnamefont {Wang}}, \bibinfo {author}
  {\bibfnamefont {S.~D.}\ \bibnamefont {Stranks}}, \bibinfo {author}
  {\bibfnamefont {H.~J.}\ \bibnamefont {Snaith}}, \ and\ \bibinfo {author}
  {\bibfnamefont {R.~J.}\ \bibnamefont {Nicholas}},\ }\href
  {http://dx.doi.org/10.1038/nphys3357} {\bibfield  {journal} {\bibinfo
  {journal} {Nat. Phys.}\ }\textbf {\bibinfo {volume} {11}},\ \bibinfo {pages}
  {582} (\bibinfo {year} {2015})}\BibitemShut {NoStop}%
\bibitem [{\citenamefont {Weller}\ \emph {et~al.}(2015)\citenamefont {Weller},
  \citenamefont {Weber}, \citenamefont {Henry}, \citenamefont {Di~Pumpo},\ and\
  \citenamefont {Hansen}}]{Weller2015}%
  \BibitemOpen
  \bibfield  {author} {\bibinfo {author} {\bibfnamefont {M.~T.}\ \bibnamefont
  {Weller}}, \bibinfo {author} {\bibfnamefont {O.~J.}\ \bibnamefont {Weber}},
  \bibinfo {author} {\bibfnamefont {P.~F.}\ \bibnamefont {Henry}}, \bibinfo
  {author} {\bibfnamefont {A.~M.}\ \bibnamefont {Di~Pumpo}}, \ and\ \bibinfo
  {author} {\bibfnamefont {T.~C.}\ \bibnamefont {Hansen}},\ }\href {\doibase
  10.1039/C4CC09944C} {\bibfield  {journal} {\bibinfo  {journal} {Chem.
  Commun.}\ }\textbf {\bibinfo {volume} {51}},\ \bibinfo {pages} {4180}
  (\bibinfo {year} {2015})}\BibitemShut {NoStop}%
\bibitem [{\citenamefont {Ren}\ \emph {et~al.}(2016)\citenamefont {Ren},
  \citenamefont {Oswald}, \citenamefont {Wang}, \citenamefont {McCandless},\
  and\ \citenamefont {Chan}}]{Ren2016}%
  \BibitemOpen
  \bibfield  {author} {\bibinfo {author} {\bibfnamefont {Y.}~\bibnamefont
  {Ren}}, \bibinfo {author} {\bibfnamefont {I.~W.~H.}\ \bibnamefont {Oswald}},
  \bibinfo {author} {\bibfnamefont {X.}~\bibnamefont {Wang}}, \bibinfo {author}
  {\bibfnamefont {G.~T.}\ \bibnamefont {McCandless}}, \ and\ \bibinfo {author}
  {\bibfnamefont {J.~Y.}\ \bibnamefont {Chan}},\ }\href {\doibase
  10.1021/acs.cgd.6b00297} {\bibfield  {journal} {\bibinfo  {journal} {Cryst.
  Growth Des.}\ }\textbf {\bibinfo {volume} {16}},\ \bibinfo {pages} {2945}
  (\bibinfo {year} {2016})}\BibitemShut {NoStop}%
\bibitem [{\citenamefont {Whitfield}\ \emph {et~al.}(2016)\citenamefont
  {Whitfield}, \citenamefont {Herron}, \citenamefont {Guise}, \citenamefont
  {Page}, \citenamefont {Cheng}, \citenamefont {Milas},\ and\ \citenamefont
  {Crawford}}]{Whitfield2016}%
  \BibitemOpen
  \bibfield  {author} {\bibinfo {author} {\bibfnamefont {P.~S.}\ \bibnamefont
  {Whitfield}}, \bibinfo {author} {\bibfnamefont {N.}~\bibnamefont {Herron}},
  \bibinfo {author} {\bibfnamefont {W.~E.}\ \bibnamefont {Guise}}, \bibinfo
  {author} {\bibfnamefont {K.}~\bibnamefont {Page}}, \bibinfo {author}
  {\bibfnamefont {Y.~Q.}\ \bibnamefont {Cheng}}, \bibinfo {author}
  {\bibfnamefont {I.}~\bibnamefont {Milas}}, \ and\ \bibinfo {author}
  {\bibfnamefont {M.~K.}\ \bibnamefont {Crawford}},\ }\href
  {http://dx.doi.org/10.1038/srep35685} {\bibfield  {journal} {\bibinfo
  {journal} {Sci. Rep.}\ }\textbf {\bibinfo {volume} {6}},\ \bibinfo {pages}
  {35685} (\bibinfo {year} {2016})}\BibitemShut {NoStop}%
\bibitem [{\citenamefont {Aryasetiawan}\ \emph {et~al.}(1996)\citenamefont
  {Aryasetiawan}, \citenamefont {Hedin},\ and\ \citenamefont
  {Karlsson}}]{Aryasetiawan1996}%
  \BibitemOpen
  \bibfield  {author} {\bibinfo {author} {\bibfnamefont {F.}~\bibnamefont
  {Aryasetiawan}}, \bibinfo {author} {\bibfnamefont {L.}~\bibnamefont {Hedin}},
  \ and\ \bibinfo {author} {\bibfnamefont {K.}~\bibnamefont {Karlsson}},\
  }\href {\doibase 10.1103/PhysRevLett.77.2268} {\bibfield  {journal} {\bibinfo
   {journal} {Phys. Rev. Lett.}\ }\textbf {\bibinfo {volume} {77}},\ \bibinfo
  {pages} {2268} (\bibinfo {year} {1996})}\BibitemShut {NoStop}%
\bibitem [{\citenamefont {Guzzo}\ \emph {et~al.}(2011)\citenamefont {Guzzo},
  \citenamefont {Lani}, \citenamefont {Sottile}, \citenamefont {Romaniello},
  \citenamefont {Gatti}, \citenamefont {Kas}, \citenamefont {Rehr},
  \citenamefont {Silly}, \citenamefont {Sirotti},\ and\ \citenamefont
  {Reining}}]{Guzzo2011}%
  \BibitemOpen
  \bibfield  {author} {\bibinfo {author} {\bibfnamefont {M.}~\bibnamefont
  {Guzzo}}, \bibinfo {author} {\bibfnamefont {G.}~\bibnamefont {Lani}},
  \bibinfo {author} {\bibfnamefont {F.}~\bibnamefont {Sottile}}, \bibinfo
  {author} {\bibfnamefont {P.}~\bibnamefont {Romaniello}}, \bibinfo {author}
  {\bibfnamefont {M.}~\bibnamefont {Gatti}}, \bibinfo {author} {\bibfnamefont
  {J.~J.}\ \bibnamefont {Kas}}, \bibinfo {author} {\bibfnamefont {J.~J.}\
  \bibnamefont {Rehr}}, \bibinfo {author} {\bibfnamefont {M.~G.}\ \bibnamefont
  {Silly}}, \bibinfo {author} {\bibfnamefont {F.}~\bibnamefont {Sirotti}}, \
  and\ \bibinfo {author} {\bibfnamefont {L.}~\bibnamefont {Reining}},\ }\href
  {\doibase 10.1103/PhysRevLett.107.166401} {\bibfield  {journal} {\bibinfo
  {journal} {Phys. Rev. Lett.}\ }\textbf {\bibinfo {volume} {107}},\ \bibinfo
  {pages} {166401} (\bibinfo {year} {2011})}\BibitemShut {NoStop}%
\bibitem [{\citenamefont {Lischner}\ \emph {et~al.}(2013)\citenamefont
  {Lischner}, \citenamefont {Vigil-Fowler},\ and\ \citenamefont
  {Louie}}]{Lischner2013}%
  \BibitemOpen
  \bibfield  {author} {\bibinfo {author} {\bibfnamefont {J.}~\bibnamefont
  {Lischner}}, \bibinfo {author} {\bibfnamefont {D.}~\bibnamefont
  {Vigil-Fowler}}, \ and\ \bibinfo {author} {\bibfnamefont {S.~G.}\
  \bibnamefont {Louie}},\ }\href {\doibase 10.1103/PhysRevLett.110.146801}
  {\bibfield  {journal} {\bibinfo  {journal} {Phys. Rev. Lett.}\ }\textbf
  {\bibinfo {volume} {110}},\ \bibinfo {pages} {146801} (\bibinfo {year}
  {2013})}\BibitemShut {NoStop}%
\bibitem [{\citenamefont {Gumhalter}\ \emph {et~al.}(2016)\citenamefont
  {Gumhalter}, \citenamefont {Kova\ifmmode~\check{c}\else \v{c}\fi{}},
  \citenamefont {Caruso}, \citenamefont {Lambert},\ and\ \citenamefont
  {Giustino}}]{gumh16}%
  \BibitemOpen
  \bibfield  {author} {\bibinfo {author} {\bibfnamefont {B.}~\bibnamefont
  {Gumhalter}}, \bibinfo {author} {\bibfnamefont {V.}~\bibnamefont
  {Kova\ifmmode~\check{c}\else \v{c}\fi{}}}, \bibinfo {author} {\bibfnamefont
  {F.}~\bibnamefont {Caruso}}, \bibinfo {author} {\bibfnamefont
  {H.}~\bibnamefont {Lambert}}, \ and\ \bibinfo {author} {\bibfnamefont
  {F.}~\bibnamefont {Giustino}},\ }\href {\doibase 10.1103/PhysRevB.94.035103}
  {\bibfield  {journal} {\bibinfo  {journal} {Phys. Rev. B}\ }\textbf {\bibinfo
  {volume} {94}},\ \bibinfo {pages} {035103} (\bibinfo {year}
  {2016})}\BibitemShut {NoStop}%
\bibitem [{\citenamefont {Verdi}\ \emph {et~al.}(2017)\citenamefont {Verdi},
  \citenamefont {Caruso},\ and\ \citenamefont {Giustino}}]{Verdi2017}%
  \BibitemOpen
  \bibfield  {author} {\bibinfo {author} {\bibfnamefont {C.}~\bibnamefont
  {Verdi}}, \bibinfo {author} {\bibfnamefont {F.}~\bibnamefont {Caruso}}, \
  and\ \bibinfo {author} {\bibfnamefont {F.}~\bibnamefont {Giustino}},\ }\href
  {http://dx.doi.org/10.1038/ncomms15769} {\bibfield  {journal} {\bibinfo
  {journal} {Nat. Commun.}\ }\textbf {\bibinfo {volume} {8}},\ \bibinfo {pages}
  {15769} (\bibinfo {year} {2017})}\BibitemShut {NoStop}%
\bibitem [{\citenamefont {Nery}\ \emph {et~al.}(2018)\citenamefont {Nery},
  \citenamefont {Allen}, \citenamefont {Antonius}, \citenamefont {Reining},
  \citenamefont {Miglio},\ and\ \citenamefont {Gonze}}]{Nery2018}%
  \BibitemOpen
  \bibfield  {author} {\bibinfo {author} {\bibfnamefont {J.~P.}\ \bibnamefont
  {Nery}}, \bibinfo {author} {\bibfnamefont {P.~B.}\ \bibnamefont {Allen}},
  \bibinfo {author} {\bibfnamefont {G.}~\bibnamefont {Antonius}}, \bibinfo
  {author} {\bibfnamefont {L.}~\bibnamefont {Reining}}, \bibinfo {author}
  {\bibfnamefont {A.}~\bibnamefont {Miglio}}, \ and\ \bibinfo {author}
  {\bibfnamefont {X.}~\bibnamefont {Gonze}},\ }\href {\doibase
  10.1103/PhysRevB.97.115145} {\bibfield  {journal} {\bibinfo  {journal} {Phys.
  Rev. B}\ }\textbf {\bibinfo {volume} {97}},\ \bibinfo {pages} {115145}
  (\bibinfo {year} {2018})}\BibitemShut {NoStop}%
\bibitem [{\citenamefont {Moser}\ \emph {et~al.}(2013)\citenamefont {Moser},
  \citenamefont {Moreschini}, \citenamefont {Ja\ifmmode \acute{c}\else
  \'{c}\fi{}imovi\ifmmode~\acute{c}\else \'{c}\fi{}}, \citenamefont
  {Bari\ifmmode \check{s}\else \v{s}\fi{}i\ifmmode~\acute{c}\else \'{c}\fi{}},
  \citenamefont {Berger}, \citenamefont {Magrez}, \citenamefont {Chang},
  \citenamefont {Kim}, \citenamefont {Bostwick}, \citenamefont {Rotenberg},
  \citenamefont {Forr\'o},\ and\ \citenamefont {Grioni}}]{Moser2013}%
  \BibitemOpen
  \bibfield  {author} {\bibinfo {author} {\bibfnamefont {S.}~\bibnamefont
  {Moser}}, \bibinfo {author} {\bibfnamefont {L.}~\bibnamefont {Moreschini}},
  \bibinfo {author} {\bibfnamefont {J.}~\bibnamefont {Ja\ifmmode \acute{c}\else
  \'{c}\fi{}imovi\ifmmode~\acute{c}\else \'{c}\fi{}}}, \bibinfo {author}
  {\bibfnamefont {O.~S.}\ \bibnamefont {Bari\ifmmode \check{s}\else
  \v{s}\fi{}i\ifmmode~\acute{c}\else \'{c}\fi{}}}, \bibinfo {author}
  {\bibfnamefont {H.}~\bibnamefont {Berger}}, \bibinfo {author} {\bibfnamefont
  {A.}~\bibnamefont {Magrez}}, \bibinfo {author} {\bibfnamefont {Y.~J.}\
  \bibnamefont {Chang}}, \bibinfo {author} {\bibfnamefont {K.~S.}\ \bibnamefont
  {Kim}}, \bibinfo {author} {\bibfnamefont {A.}~\bibnamefont {Bostwick}},
  \bibinfo {author} {\bibfnamefont {E.}~\bibnamefont {Rotenberg}}, \bibinfo
  {author} {\bibfnamefont {L.}~\bibnamefont {Forr\'o}}, \ and\ \bibinfo
  {author} {\bibfnamefont {M.}~\bibnamefont {Grioni}},\ }\href {\doibase
  10.1103/PhysRevLett.110.196403} {\bibfield  {journal} {\bibinfo  {journal}
  {Phys. Rev. Lett.}\ }\textbf {\bibinfo {volume} {110}},\ \bibinfo {pages}
  {196403} (\bibinfo {year} {2013})}\BibitemShut {NoStop}%
\bibitem [{\citenamefont {Chen}\ \emph {et~al.}(2015)\citenamefont {Chen},
  \citenamefont {Avila}, \citenamefont {Frantzeskakis}, \citenamefont {Levy},\
  and\ \citenamefont {Asensio}}]{Chen2015a}%
  \BibitemOpen
  \bibfield  {author} {\bibinfo {author} {\bibfnamefont {C.}~\bibnamefont
  {Chen}}, \bibinfo {author} {\bibfnamefont {J.}~\bibnamefont {Avila}},
  \bibinfo {author} {\bibfnamefont {E.}~\bibnamefont {Frantzeskakis}}, \bibinfo
  {author} {\bibfnamefont {A.}~\bibnamefont {Levy}}, \ and\ \bibinfo {author}
  {\bibfnamefont {M.~C.}\ \bibnamefont {Asensio}},\ }\href
  {http://dx.doi.org/10.1038/ncomms9585} {\bibfield  {journal} {\bibinfo
  {journal} {Nat. Commun.}\ }\textbf {\bibinfo {volume} {6}},\ \bibinfo {pages}
  {8585} (\bibinfo {year} {2015})}\BibitemShut {NoStop}%
\bibitem [{\citenamefont {Cancellieri}\ \emph {et~al.}(2016)\citenamefont
  {Cancellieri}, \citenamefont {Mishchenko}, \citenamefont {Aschauer},
  \citenamefont {Filippetti}, \citenamefont {Faber}, \citenamefont
  {Bari\v{s}i\'c}, \citenamefont {Rogalev}, \citenamefont {Schmitt},
  \citenamefont {Nagaosa},\ and\ \citenamefont {Strocov}}]{Cancellieri2016}%
  \BibitemOpen
  \bibfield  {author} {\bibinfo {author} {\bibfnamefont {C.}~\bibnamefont
  {Cancellieri}}, \bibinfo {author} {\bibfnamefont {A.~S.}\ \bibnamefont
  {Mishchenko}}, \bibinfo {author} {\bibfnamefont {U.}~\bibnamefont
  {Aschauer}}, \bibinfo {author} {\bibfnamefont {A.}~\bibnamefont
  {Filippetti}}, \bibinfo {author} {\bibfnamefont {C.}~\bibnamefont {Faber}},
  \bibinfo {author} {\bibfnamefont {O.~S.}\ \bibnamefont {Bari\v{s}i\'c}},
  \bibinfo {author} {\bibfnamefont {V.~A.}\ \bibnamefont {Rogalev}}, \bibinfo
  {author} {\bibfnamefont {T.}~\bibnamefont {Schmitt}}, \bibinfo {author}
  {\bibfnamefont {N.}~\bibnamefont {Nagaosa}}, \ and\ \bibinfo {author}
  {\bibfnamefont {V.~N.}\ \bibnamefont {Strocov}},\ }\href
  {http://dx.doi.org/10.1038/ncomms10386} {\bibfield  {journal} {\bibinfo
  {journal} {Nat. Commun.}\ }\textbf {\bibinfo {volume} {7}},\ \bibinfo {pages}
  {10386} (\bibinfo {year} {2016})}\BibitemShut {NoStop}%
\bibitem [{\citenamefont {Wang}\ \emph {et~al.}(2016)\citenamefont {Wang},
  \citenamefont {McKeown~Walker}, \citenamefont {Tamai}, \citenamefont {Wang},
  \citenamefont {Ristic}, \citenamefont {Bruno}, \citenamefont {de~la Torre},
  \citenamefont {Ricc\`o}, \citenamefont {Plumb}, \citenamefont {Shi},
  \citenamefont {Hlawenka}, \citenamefont {S\'anchez-Barriga}, \citenamefont
  {Varykhalov}, \citenamefont {Kim}, \citenamefont {Hoesch}, \citenamefont
  {King}, \citenamefont {Meevasana}, \citenamefont {Diebold}, \citenamefont
  {Mesot}, \citenamefont {Moritz}, \citenamefont {Devereaux}, \citenamefont
  {Radovic},\ and\ \citenamefont {Baumberger}}]{Wang2016}%
  \BibitemOpen
  \bibfield  {author} {\bibinfo {author} {\bibfnamefont {Z.}~\bibnamefont
  {Wang}}, \bibinfo {author} {\bibfnamefont {S.}~\bibnamefont
  {McKeown~Walker}}, \bibinfo {author} {\bibfnamefont {A.}~\bibnamefont
  {Tamai}}, \bibinfo {author} {\bibfnamefont {Y.}~\bibnamefont {Wang}},
  \bibinfo {author} {\bibfnamefont {Z.}~\bibnamefont {Ristic}}, \bibinfo
  {author} {\bibfnamefont {F.~Y.}\ \bibnamefont {Bruno}}, \bibinfo {author}
  {\bibfnamefont {A.}~\bibnamefont {de~la Torre}}, \bibinfo {author}
  {\bibfnamefont {S.}~\bibnamefont {Ricc\`o}}, \bibinfo {author} {\bibfnamefont
  {N.~C.}\ \bibnamefont {Plumb}}, \bibinfo {author} {\bibfnamefont
  {M.}~\bibnamefont {Shi}}, \bibinfo {author} {\bibfnamefont {P.}~\bibnamefont
  {Hlawenka}}, \bibinfo {author} {\bibfnamefont {J.}~\bibnamefont
  {S\'anchez-Barriga}}, \bibinfo {author} {\bibfnamefont {A.}~\bibnamefont
  {Varykhalov}}, \bibinfo {author} {\bibfnamefont {T.~K.}\ \bibnamefont {Kim}},
  \bibinfo {author} {\bibfnamefont {M.}~\bibnamefont {Hoesch}}, \bibinfo
  {author} {\bibfnamefont {P.~D.~C.}\ \bibnamefont {King}}, \bibinfo {author}
  {\bibfnamefont {W.}~\bibnamefont {Meevasana}}, \bibinfo {author}
  {\bibfnamefont {U.}~\bibnamefont {Diebold}}, \bibinfo {author} {\bibfnamefont
  {J.}~\bibnamefont {Mesot}}, \bibinfo {author} {\bibfnamefont
  {B.}~\bibnamefont {Moritz}}, \bibinfo {author} {\bibfnamefont {T.~P.}\
  \bibnamefont {Devereaux}}, \bibinfo {author} {\bibfnamefont {M.}~\bibnamefont
  {Radovic}}, \ and\ \bibinfo {author} {\bibfnamefont {F.}~\bibnamefont
  {Baumberger}},\ }\href {http://dx.doi.org/10.1038/nmat4623} {\bibfield
  {journal} {\bibinfo  {journal} {Nat. Mater.}\ }\textbf {\bibinfo {volume}
  {15}},\ \bibinfo {pages} {835} (\bibinfo {year} {2016})}\BibitemShut
  {NoStop}%
\bibitem [{\citenamefont {Feynman}(1955)}]{Feynman1955}%
  \BibitemOpen
  \bibfield  {author} {\bibinfo {author} {\bibfnamefont {R.~P.}\ \bibnamefont
  {Feynman}},\ }\href {\doibase 10.1103/PhysRev.97.660} {\bibfield  {journal}
  {\bibinfo  {journal} {Phys. Rev.}\ }\textbf {\bibinfo {volume} {97}},\
  \bibinfo {pages} {660} (\bibinfo {year} {1955})}\BibitemShut {NoStop}%
\bibitem [{\citenamefont {Schultz}(1959)}]{Schultz1959}%
  \BibitemOpen
  \bibfield  {author} {\bibinfo {author} {\bibfnamefont {T.~D.}\ \bibnamefont
  {Schultz}},\ }\href {\doibase 10.1103/PhysRev.116.526} {\bibfield  {journal}
  {\bibinfo  {journal} {Phys. Rev.}\ }\textbf {\bibinfo {volume} {116}},\
  \bibinfo {pages} {526} (\bibinfo {year} {1959})}\BibitemShut {NoStop}%
\bibitem [{\citenamefont {Davies}\ \emph {et~al.}(2018)\citenamefont {Davies},
  \citenamefont {Filip}, \citenamefont {Patel}, \citenamefont {Crothers},
  \citenamefont {Verdi}, \citenamefont {Wright}, \citenamefont {Milot},
  \citenamefont {Giustino}, \citenamefont {Johnston},\ and\ \citenamefont
  {Herz}}]{Davies2018}%
  \BibitemOpen
  \bibfield  {author} {\bibinfo {author} {\bibfnamefont {C.~L.}\ \bibnamefont
  {Davies}}, \bibinfo {author} {\bibfnamefont {M.~R.}\ \bibnamefont {Filip}},
  \bibinfo {author} {\bibfnamefont {J.~B.}\ \bibnamefont {Patel}}, \bibinfo
  {author} {\bibfnamefont {T.~W.}\ \bibnamefont {Crothers}}, \bibinfo {author}
  {\bibfnamefont {C.}~\bibnamefont {Verdi}}, \bibinfo {author} {\bibfnamefont
  {A.~D.}\ \bibnamefont {Wright}}, \bibinfo {author} {\bibfnamefont {R.~L.}\
  \bibnamefont {Milot}}, \bibinfo {author} {\bibfnamefont {F.}~\bibnamefont
  {Giustino}}, \bibinfo {author} {\bibfnamefont {M.~B.}\ \bibnamefont
  {Johnston}}, \ and\ \bibinfo {author} {\bibfnamefont {L.~M.}\ \bibnamefont
  {Herz}},\ }\href {https://doi.org/10.1038/s41467-017-02670-2} {\bibfield
  {journal} {\bibinfo  {journal} {Nat. Commun.}\ }\textbf {\bibinfo {volume}
  {9}},\ \bibinfo {pages} {293} (\bibinfo {year} {2018})}\BibitemShut {NoStop}%
\bibitem [{\citenamefont {Momma}\ and\ \citenamefont {Izumi}(2008)}]{vesta}%
  \BibitemOpen
  \bibfield  {author} {\bibinfo {author} {\bibfnamefont {K.}~\bibnamefont
  {Momma}}\ and\ \bibinfo {author} {\bibfnamefont {F.}~\bibnamefont {Izumi}},\
  }\href {\doibase 10.1107/S0021889808012016} {\bibfield  {journal} {\bibinfo
  {journal} {J. Appl. Crystallogr.}\ }\textbf {\bibinfo {volume} {41}},\
  \bibinfo {pages} {653} (\bibinfo {year} {2008})}\BibitemShut {NoStop}%
\end{thebibliography}%

\end{document}